\documentclass[fontsize=11pt,a4paper,toc=flat]{arxiv-article}
\usepackage{standalone}
\usepackage{Definitions}
\usepackage{picture-things}
\usetikzlibrary{decorations.markings} %

\newcommand{\cO}{\mathcal{O}}

\usepackage{comment}

\usepackage{hyperref}
\graphicspath{{./plots/}{./FIGS/}}

\allowdisplaybreaks
\newcommand{\OfficialTitle}{Semiclassics at the cusp
}
\title{\setstretch{1.4}
	{\color{Thoughtless}\textls[-20]{\OfficialTitle}}
}

\hypersetup{pdfauthor={Jahmall Bersini, Domenico Orlando, Susanne Reffert, Jesse Woods},pdftitle={\OfficialTitle},%
	colorlinks=true,linkcolor=ThoughtYouWere,citecolor=ThoughtYouWere,urlcolor=ThoughtYouWere}

\author{%
	\begin{minipage}{.94\textwidth}
		\begin{center} \dosserif%
			{\small
           \textbf{Jahmall Bersini}\textsuperscript{\ding{73}},
           \textbf{Domenico~Orlando}\textsuperscript{\ding{72}\ding{73}}
         	\textbf{Susanne~Reffert}\textsuperscript{\ding{73}}, and
         	\textbf{Jesse Woods}\textsuperscript{\ding{73}}  			
			}
		\end{center}
		\authorBlock{\ding{73}}{\dosserif{}%
			Albert Einstein Center for Fundamental Physics,\\
			Institute for Theoretical Physics, University of Bern,\\
			Sidlerstrasse 5, CH-3012 Bern, Switzerland}
		\authorBlock{\ding{72}}{\dosserif{}%
			INFN sezione di Torino.\\
			via Pietro Giuria 1, 10125 Torino, Italy}
	\end{minipage}
}

\date{}

\begin{document}

\numberwithin{equation}{section}

\begin{titlepage}

\maketitle

	\thispagestyle{empty}

   \vfill

   \abstract{\normalfont{}\noindent{}%
     We study cusped Wilson line operators in the Abelian Higgs model in \( d = 4 - \epsilon \) at large external charges.
     Using a double-scaling limit \( Q \to \infty \), \( \epsilon \to 0 \) with \( Q\epsilon \) fixed, we develop a semiclassical framework that provides analytic control beyond fixed-order perturbation theory.
     We compute the cusp anomalous dimension for arbitrary charges up to next-to-next-to-leading order in the gauge coupling, while resumming scalar self-interactions to all orders.
     Our results interpolate between perturbative and large-charge regimes, accessing domains that are invisible in fixed-order perturbation theory.
     As an application, we provide new predictions for various defect \acs{cft} observables, including the Mandelstam–Schwinger-dressed two-point function characterizing the superconducting phase transition.
   }

\end{titlepage}

\setstretch{1.1}
\tableofcontents

\newpage
\section{Introduction}
\label{sec:intro}

Semiclassical calculations in a regime of large charge, or in a double-scaling limit involving large charge, have proven to be an extremely successful tool for accessing strongly coupled systems over the last decade~\cite{Hellerman:2015nra,Gaume:2020bmp}. This framework has been applied to the three-dimensional \(O(N)\) model at the Wilson--Fisher point~\cite{Alvarez-Gaume:2016vff, Alvarez-Gaume:2019biu, Dondi:2021buw} and has subsequently been extended to fermionic models~\cite{Dondi:2022zna, Antipin:2022naw, Bersini:2025lxs}.
An obvious question is how to apply this formalism to systems with gauge symmetry. This immediately raises a variety of issues related to the requirement that observables be gauge invariant. The Gauss constraint implies that gauge-invariant charged operators must be non-local.
It follows that, by Elitzur's theorem~\cite{Elitzur:1975im}, the naive \(n\)-point correlation function of charged fields vanishes identically.
The traditional way to generalize these correlators to gauge theories has been to dress products of operators \emph{à la} Dirac~\cite{Dirac:1955uv} by including the associated Coulomb fields%
\footnote{The Dirac case has been studied at large-charge in~\cite{Antipin:2022hfe}, and we revisit these results in Appendix~\ref{sec:Dirac}.}%
, or \emph{à la} Mandelstam and Schwinger~\cite{Mandelstam:1962mi,Schwinger:1959xd,Schwinger:1962tn,Schwinger:1962tp} by adding a Wilson line joining the insertions. 

In this article, we take a general approach and consider a class of observables constructed from line intersections. The simplest case is a cusp: two semi-infinite lines carrying different charges \(Q_1\) and \(Q_2\) meeting at a point (the cusp).
The problem of line operators in \acp{cft} has been addressed repeatedly in the recent literature~\cite{Cuomo:2021rkm, Rodriguez-Gomez:2022gbz, Bianchi:2022sbz, Gimenez-Grau:2022czc, Aharony:2022ntz, Aharony:2023amq, Cuomo:2024psk,Iqbal:2025yop, Giombi:2025evu, DAlise:2025knv}. 
The so-called cusp anomalous dimension \(\Gamma_{Q_1 Q_2}\) is the primary quantity of interest, along with the spectrum. It measures the strength of the divergence produced by a cusp in a Wilson loop.
The expectation value of these cusp operators is logarithmically divergent,
\begin{equation}
  \log\Braket{\littlecusp} = -\Gamma_{Q_1 Q_2}(\alpha_{*}) \log\frac{L}{a} + \dots,
\end{equation}
where \(\alpha_{*}\) is the angle between the two arms, $L$ is the length of the line, and $a$ is an \ac{uv} regulator, which can be thought of as the microscopic scale at which the cusp is smoothed.
The dots stand for non-logarithmic divergences, such as power-law ``cosmological constant'' contributions proportional to $L/a$.%

Given that one can think of a Wilson line as the trajectory of a heavy particle, these cusps represent sudden changes in velocity and thus yield information about Bremsstrahlung:
the sudden acceleration forces the gauge field to radiate, and the probability amplitude for this radiation is singular in both the \ac{uv} and \ac{ir} limits.
The cusp anomalous dimension characterizes the \ac{ir} divergences of scattering massive colored particles~\cite{Beisert:2006ez} and is identified with the energy of a static quark potential on the cylinder~\cite{Correa:2012hh,Bianchi:2018zpb,Bianchi:2019dlw}.
The properties of such cusps for large angles have been studied in~\cite{Korchemsky:1987wg}.
It has also been shown that these cusps coincide with heavy quark currents in heavy quark effective field theory~\cite{Korchemsky:1991zp}.
In the limit of a backtracking cusp, $\alpha_{*} \to 0$, the anomalous dimension yields information about the static quark-antiquark potential.
Perturbative calculations were computed to two loops for $\mathcal{N}=4$ \ac{sym} in~\cite{Makeenko:2006ds} in the 't Hooft planar limit.
These results have been extended to three loops in \ac{qcd} and its supersymmetric extensions in~\cite{Grozin:2015kna} and to four loops for small angles in~\cite{Bruser:2019auj} and for rectangular Wilson loops in~\cite{Henn:2019swt}.
At five loops, the Abelian color structures are known, as well as all-order contributions for terms pertaining to the conformal anomaly~\cite{Grozin:2022umo}.
In the supersymmetric case one can also use integrability and the cusp anomalous dimension is in principle known from the spectral curve~\cite{Gromov:2015dfa,Cavaglia:2018lxi}.

A relatively modern perspective on Wilson lines and other extended operators is in terms of defect operators~\cite{Kapustin:2005py,Drukker:2010jp}: their insertion in the path integral can be absorbed as a new term in the action localized on a sub-manifold.
Defects have been widely studied due to the relevance of impurities or boundaries~\cite{kondo1970theory,Wilson:1974mb,Billo:2013jda,Cuomo:2021cnb,Cuomo:2022xgw}, as well as brane dynamics~\cite{Constable:2002xt}.
At criticality, one is led to study defect \ac{cft}~\cite{Cardy:1984bb,Cardy:1991tv}, which has additional \ac{cft} data encoded in bulk to boundary correlators~\cite{McAvity:1993ue,McAvity:1995zd,Gliozzi:2015qsa,Billo:2016cpy}.
In defect \ac{cft}, a simple example arises at \(\alpha_{*}= \pi \) (no deflection).
For $Q_1\neq Q_2$ along the two half-lines, the cusp anomalous dimension is the scaling dimension $\Delta_{Q_1 Q_2}$ of the lowest defect-changing operator, which may be viewed as a local operator $\cO_{Q_1 Q_2}$ placed at the junction of two arms~\cite{affleck1994fermi,Affleck:1996mm,Cuomo:2024psk},
\begin{equation}
  \Gamma_{Q_1Q_2}(\pi) = \Delta_{Q_1 Q_2}\, .
\end{equation} 
As discussed in~\cite{Lanzetta:2025xfw}, the spectrum of defect-changing operators determines the stability of line defects that spontaneously break global symmetries.
For \(Q_2 = 0\), the dimension of the defect-creation operator $\Delta_{Q 0}$ controls the critical behavior of the \ac{ms} dressed two-point function, which is of interest in characterizing the superconducting phase transition~\cite{Kleinert_2003, Kiometzis:1995eg, Herbut}.

\bigskip

A natural setting for studying the behavior of a cusp in a gauge \ac{cft} is the Abelian Higgs model.
At weak coupling, this is a system of \(N\) scalar fields, all with unit charge under a gauged \(U(1)\) symmetry.
The theory is known to flow in the \ac{ir} to an interacting \ac{cft}.
In \( 4 - \epsilon\) dimensions, the fixed point is under perturbative control and is real for \(N\) sufficiently large~\cite{Halperin:1973jh, Herbut, Ihrig:2019kfv}.
At criticality, it describes phase transitions in a variety of physical systems ranging from superconductors~\cite{Landau:1937obd,Herbut, Dasgupta:1981zz} to liquid crystals~\cite{Halperin:1973jh} and cosmic strings~\cite{Hindmarsh:2008dw}.
The cusp is identified with two semi-infinite straight Wilson lines meeting at the origin, where a scalar field is inserted to satisfy the Gauss constraint (the presence of matter in the theory makes it possible to end Wilson lines).

In this paper, we use a semiclassical approach to calculate the quantities of interest, such as the cusp anomalous dimension and the spectrum, concentrating on the limit where both charges are large and of the the same order \(Q \gg 1\).
In this regime, we use a double-scaling limit \(Q \to \infty\), \(\epsilon \to 0\) with \(Q \epsilon\) fixed in which the path integral localizes around the saddles of a reduced action that takes into account the presence of the cusp.
This action is written in terms of appropriate 't Hooft-like couplings, \(\kappa\) for the scalar self-interactions and \(G\) for the gauge coupling.
Its quantum fluctuations are controlled by \(\hbar = 1/ Q\).
We obtain a controlled expansion in inverse powers of the charge, even when the microscopic couplings are not parametrically small.
Working at large charge turns the Wilson line into a semiclassical object whose backreaction on the gauge and scalar fields can be treated systematically.

In spite of the resulting major simplifications, the full solution to the \acl{eom} remains elusive because we are dealing with a set of non-linear \acp{pde}.
We therefore limit ourselves here to a regime of small \(G\) with \(\kappa\) fixed, in which the semiclassical analysis partially overlaps with the standard perturbative approach.
This allows us both to validate our technique and to extend previous results found in the literature~\cite{Kleinert:2005sa, Kleinert_2003}.
The main advantage of our approach is that we can access regimes that are invisible in fixed-order perturbation theory, where interactions are effectively strong due to large sources, despite weak microscopic couplings.
Other regimes will be explored in a companion paper.
Our main result is the expression for the cusp anomalous dimension at \ac{nnlo} in \(G\):
\begin{multline}
   \Gamma_{q_1 q_2}(\alpha_{*}) = Q \Bigg[ \Gamma_{q_1 q_2}^{(-1, 0)} - \frac{3 \left(q_1 - q_2 \right)^2 + 4 q_1 q_2 \left( 1 + \left( \pi - \alpha_{*} \right) \cot(\alpha_{*}) \right)}{16 \pi^2} G \\
  + \Gamma_{q_1 q_2}^{(-1,2)}( \alpha_{*}) G^2  + \order{G^3} \Bigg] + \Gamma_{q_1 q_2}^{(0,0)} + \order{Q^0 G,1/Q}\,,
\end{multline}
where $q_{1,2}=Q_{1,2}/Q$, and \(\Gamma^{(-1,0)}\), \(\Gamma^{(-1,2)}\) and \(\Gamma^{(0,0)}\) are rather cumbersome but exact functions of the quartic coupling \(\kappa\) which resum the contribution of an infinite series of Feynman diagrams.
The \(\Gamma^{(0,0)}\) contribution comes from the fluctuations over the large-charge saddle for $G=0$. For finite but small $G$, we show that, because of the Higgs mechanism, there is no massless type-I Goldstone mode in the spectrum and the gauge field acquires a mass unless \(Q_1 = Q_2\).
This is important because in the standard large-charge \ac{eft} for systems with global symmetry, the spectrum of the Goldstone contains the \ac{cft} descendants, which here are missing.

In the small-$\kappa$ regime, where our results reproduce known one-loop results \cite{Kleinert_2003, Antipin:2022hfe} and extend them to higher orders, we obtain
\begin{multline}
  \Gamma_{q_1 q_2}(\alpha_{*}) =
    Q \Bigg[\pqty*{\pqty*{q_1 - q_2}+\frac{\pqty*{q_1 - q_2}^2 \kappa }{8 \pi ^2}-\frac{\pqty*{q_1 - q_2}^3 \kappa ^2}{32 \pi ^4}+ \dots} 
    \\
    - \frac{3 \left(q_1 - q_2 \right)^2 + 4 q_1 q_2  \left( 1 + \left( \pi - \alpha_{*} \right) \cot(\alpha_{*}) \right)}{16 \pi^2} G \\
    +  \Bigg( \tfrac{q_1-q_2}{384 \pi ^4} \left(8 q_1 q_2 \left(\alpha_* (\alpha_*-2 \pi ) (\alpha_*-\pi ) \cot (\alpha_*)-2 \pi ^2\right) +\left(3-8 \pi ^2\right) (q_1-q_2)^2\right) \\
    + \tfrac{(q_1-q_2)^2}{5760 \pi ^6} \left(15 \left(\left(\alpha_*^2-12\right) \alpha_*^2-4 \pi  \left(\alpha_*^2-6\right) \alpha_*+4 \pi ^2 \left(\alpha_*^2-3\right)\right. \right. \\
    \left. \left. +6 (\alpha_*-2 \pi ) (\alpha_*-\pi ) \alpha_* \cot (\alpha_*)\right)q_1 q_2 +\left(90-30 \pi ^2+4 \pi ^4\right) (q_1-q_2)^2\right) \kappa\\
    + \tfrac{(q_1-q_2)^3 }{184320 \pi ^8}\left(\left(765-720 \pi ^2+32 \pi ^4\right) (q_1-q_2)^2  \right. \\
    \left.  +8 q_1 q_2 \left( -16 \pi ^4 +45 \left(\alpha_*^2-1\right) \alpha_*^2+90 \pi  \left(1-2 \alpha_*^2\right) \alpha_*  +30 \pi ^2 \left(6 \alpha_*^2-7\right)  \right. \right. \\
    \left. \left.-(\alpha_*-2 \pi ) (\alpha_*-\pi ) \left(6 \alpha_*^2-12 \pi  \alpha_*-8 \pi ^2-105\right) \alpha_* \cot \alpha_*\right) \right) \kappa^2
    + \dots \Bigg) G^2 + \order{G^3} \Bigg]\\
    -\left[\tfrac{(N+5) (q_1-q_2)\kappa}{8 \pi ^2}+\tfrac{(N-3) (q_1-q_2)^2\kappa ^2 }{64 \pi ^4}+ 
      \dots\right] +\order{Q^0 G} \,.
\end{multline}
In the opposite strongly-coupled $\kappa \gg 1$ limit, which was previously inaccessible, the cusp anomalous dimension reads
 \begin{multline}
  \Gamma_{q_1 q_2}(\alpha_{*}) =
    Q \Bigg[\pqty*{\tfrac{3}{4}\left(\tfrac{\kappa  (q_1-q_2)^4}{2 \pi ^2}\right)^{1/3}+\dots} 
   - \frac{3 \left(q_1 - q_2 \right)^2 + 4 q_1 q_2  \left( 1 + \left( \pi - \alpha_{*} \right) \cot(\alpha_{*}) \right)}{16 \pi^2} G  \\
      -  \pqty*{\tfrac{1}{64}\left(\left(3+4 \pi ^2\right) (q_1-q_2)^2+12 q_1 q_2(2\pi-\alpha_*)\alpha_*\right) \left(\tfrac{(q_1-q_2)^2}{4 \pi ^{10} \kappa }\right)^{1/3} +\dots}G^2 + \dots \Bigg]  \\
      +\bigg[\tfrac{1}{192}\Big(4 (N+4) \log ( (q_1-q_2)\kappa ) -4 (3 \sqrt{3}+13+8 \log (4 \pi )-15 \operatorname{arccoth}(\sqrt{3})) \\
      -N (15+8 \log (4 \pi ))  +12 \gamma_E  (N+4) \Big)\left(\tfrac{\kappa ^4 (q_1-q_2)^4}{2 \pi ^8}\right)^{1/3}+ \dots\bigg] +\order{Q^0 G} \,,
\end{multline}
where $\gamma_E$ is Euler's constant. 

These expressions contain, as special cases, the defect-changing operator when the two arms are parallel, \(\alpha_{*} = \pi\), and a \emph{half-Wilson line} for zero aperture \(\alpha_{*} = 0 \).
In this latter case, we can extract the scaling of the \ac{ms}-dressed two-point function and show that the
previously conjectured equivalence with a scalar two-point function in traceless gauge~\cite{Kleinert_2003,Kleinert:2005sa} breaks down beyond leading order in \(G\).

Another interesting observation is that we have enough control over the perturbative expansion in the gauge coupling to show that it breaks down for \(G \approx 2 \pi\), signaling a possible phase transition, in analogy to the one found for straight Wilson lines~\cite{Aharony:2022ntz}.

\bigskip

The plan of this paper is as follows.
In Section~\refstring{sec:setup}, we describe our setup, starting with the actions and defining the Wilson line with a cusp.
The main objective of this work is to compute the cusp anomalous dimension. We set out to compute it semiclassically in Section~\refstring{sec:semiclassical-cusp}, discussing first the \ac{eom} and the boundary contributions. Despite the double-scaling limit, we still need to work perturbatively in the gauge coupling $G$. We give the contributions to the cusp anomalous dimension up to \ac{nnlo} in $G$.
In Section~\refstring{symmental}, we discuss the symmetries and the resulting spectrum.
Once we have understood the fluctuations, we can put our results together to find the full cusp anomalous dimension in Section~\refstring{sec:cusp-dimension}.
We also discuss the relevant limiting cases in this section.
In Section~\refstring{sec:pert-epsil-expans}, we match our semi-classical result to the standard perturbative calculation.
In Section~\refstring{sec:conclusions}, we give conclusions and outlook.
Appendix~\ref{sec:Dirac} revisits the Dirac dressing. In particular, we show the validity of the state-operator correspondence for the dressed field operator despite the latter being nonlocal.
In Appendix~\ref{sec:explicit} we collect explicit expressions for the various contributions to $\Gamma_{q_1 q_2}(\alpha_{*})$.
Finally, in Appendix~\ref{spettro} we present the spectrum of the theory to linear order in the gauge coupling $G$.

\section{Setup}
\label{sec:setup}

The Abelian Higgs model describes the dynamics of \(N\) complex scalar fields with unit charge under a gauged \(U(1)\) symmetry.
In Euclidean signature, the action is given by
\begin{equation}
  S_{\text{AH}}[A, \phi] = \int \dd{x} \left[\frac{1}{4} F_{\mu\nu} F^{\mu\nu} + (D^{\mu} \phi_a)^{*} D_{\mu} \phi_{a} + \frac{(4 \pi)^2}{6}\lambda (\phi_a^{*} \phi_a)^2 \right] \, ,
\end{equation}
with \(D_{\mu} = \del_{\mu}{} + i e A_{\mu}\). 
The global symmetry of the theory is $PSU(N)=SU(N)/\mathbb{Z}_N$, where the $\mathbb{Z}_N$ factor arises because a transformation in the center of $SU(N)$ can be undone by $U(1)$ gauge transformations that act as $\phi\to e^{i e \Lambda(x)}\phi$ and $A_{\mu}\to
A_{\mu}-\del_{\mu}\Lambda(x)$. It is well known that in \(d = 4 - \epsilon\) the model has a perturbative fixed point for \(\epsilon \to 0\), where both couplings \(\lambda\) and \(e^2\) are under perturbative control~\cite{Halperin:1973jh}
\begin{align}
  \lambda^* = \frac{3\left( N +18 \pm \sqrt{(N-180) N-540}\right) }{4 N(N+4)}  \epsilon + \order{\epsilon^2} \,, && \pqty{e^{*}}^2 = \frac{24 \pi ^2 }{N}  \epsilon + \order{\epsilon^2}  \, .
\end{align}
At this order, the fixed points are real for $N \ge 183$.
Estimates based on state-of-the-art four-loop calculations~\cite{Ihrig:2019kfv} reduce the critical value down to $N \approx 12$.  

At weak coupling we can identify a cusp with two Wilson lines of charge \(Q_1 > Q_2\) ending on the insertion of a charged scalar operator
at the origin:%
\begin{equation}
  Z_{Q_1 Q_2}(\alpha_{*}) = \Braket{\littlecusp} = \braket{\cO_{\substack{\yngrow{4}\\Q_1-Q_2}}^{*}(x_\infty) \,W_{\gamma_1}(Q_1) \, \cO_{\substack{\yngrow{4}\\Q_1-Q_2}}(0) \, W_{\gamma_2}(-Q_2)} \,, 
\end{equation}
where \(\gamma_1\) and \(\gamma_2\) are straight lines from the origin to infinity, with relative angle \(\alpha_{*}\).\footnote{In the literature, the cusp angle often refers to the deflection angle $\varphi=\pi-\alpha_*$.}
For later purposes we can write the line insertion as
\begin{equation}
  W_{\gamma}(Q) = e^{i e Q \int_{\gamma} A} = e^{i e Q \int J_{\text{MS}}^{\mu}(x \mid \gamma) A_\mu(x) \dd{x}}  \,. 
\end{equation}
where the current \(J_{\text{MS}}^{\mu}\) defines the so-called \ac{ms} dressing~\cite{Mandelstam:1962mi,Schwinger:1959xd,Schwinger:1962tn,Schwinger:1962tp}
and is a delta function supported along the straight line $\gamma$ parametrised by $s\in[0,1]$
\begin{equation}
  \label{eq:JMS}
    J_{\text{MS}}^\mu(x \mid \gamma)=\int_0^1 \dd{s} \delta\left(x-\gamma(s)\right)\frac{d\gamma^\mu(s)}{ds}.
\end{equation}
By gauge invariance, the defect-changing operator \(\cO(0)\) needs to have \(U(1)\) charge \(Q_1 - Q_2\).
In fact, gauge invariance requires the insertion of a second operator \(\cO^{*}(x_{\infty})\) with opposite charge at infinity.
In different conformal frames, either insertion can be treated as a boundary condition.

As for the global symmetry, we choose the cusp to transform in the completely symmetric representation \(\operatorname{Sym}^{Q_1-Q_2}\) of $SU(N)$.
In the perturbative regime, we can identify the defect-changing operator with the symmetrized product (indices can be repeated)
\begin{equation}
  \cO_{\substack{\yngrow{4}\\Q_1-Q_2}}(x) = \phi_{( i_1} (x) \dots \phi_{ i_{Q_1-Q_2})}(x) \, .
\end{equation}
This is expected to be the lowest defect-changing operator since other representations would require adding derivatives to realize an antisymmetrization, which generically increases the energy of the corresponding semiclassical state.%
\footnote{For $Q_1-Q_2 \mod N \neq 0$ the defect-changing operator transforms according to a projective representation of $PSU(N)$.}

To perform the semiclassical calculation, we can pick any representative of the irreducible representation: without loss of generality we will take  $\cO(x) = \phi_1^{Q_1-Q_2}(x)$ and look for a classical profile where $\phi_{\ge2} = 0$.
For ease of notation we will drop the subscript \(\phi \equiv \phi_1\) when this does not generate confusion.

\bigskip

Since there is a perturbative fixed point, we will study the system in a double-scaling limit where both charges \(Q_1\) and \(Q_2\) are large, and the product \(Q \epsilon\) is kept fixed.
The advantage of such a limit is that we will be able to study the system semiclassically.
To see this, we need to show that the path integral localizes around its saddle and the quantum corrections are controlled by \(1/Q\).

First, we introduce the rescaled order-$1$ couplings and charges:
\begin{align}
  q_1 &= \frac{Q_1}{Q} \, , & q_2 &= \frac{Q_2}{Q} \, , &   G &= Q e^2 , & \kappa &=  \frac{(4 \pi)^2}{6}Q \lambda\ ,
\end{align}
where \(Q\) is some large number such that \(q_1\) and \(q_2\) are parametrically of order \(\order{Q^0}\).%
 
 Next, we collect all the terms in the integrand into a single exponential:
\begin{equation}
  \Braket{\littlecusp} = \int \DD{A} \DD{\phi_a} e^{- S_{AH}[A, \phi] - S_{\text{ins}}[A, \phi]},
\end{equation}
where the insertion term is given by
\begin{equation}
  S_{\text{ins}}[A, \phi] = \left(Q_2 - Q_1\right) \int \dd{x} \pqty*{\log( \phi_1 ) \delta(x) + \log( \phi^{*}_1) \delta(x - x_{\infty}) }- i e Q_1 \int_{\gamma_1} A + i e Q_2 \int_{\gamma_2} A \,.
\end{equation}
Rescaling the fields as
\begin{align}
  \phi & \to \sqrt{Q} \phi, & A &\to \sqrt{Q} A \,, 
\end{align}
we find
\begin{multline}
  \label{eq:cusp-rescaled-action}
  S_{\text{AH}} + S_{\text{ins}} = Q \bar S  = Q \Bigg[ \int \dd{x} \Big( \frac{1}{4} F_{\mu\nu} F^{\mu\nu} + \pqty*{\del_{\mu} + i \sqrt{G} A_{\mu}} \phi_a^{*} \pqty*{\del_{\mu} - i \sqrt{G} A_{\mu}} \phi_a + \kappa \pqty*{\phi_a^*\phi_a}^2\\
  + \left( q_2 - q_1 \right) \pqty*{\log( \phi_1 ) \delta(x) + \log( \phi^{*}_1) \delta(x - x_{\infty}) } \Big) \\ - i q_1 \sqrt{G} \int_{\gamma_1} A + i q_2 \sqrt{G} \int_{\gamma_2} A \Bigg] \, ,
\end{multline}
showing manifestly that \(Q\) plays the role of \(1/\hbar\) for the semiclassical expansion around the saddle of \(\bar S\).
Note that at the leading semiclassical order, one can neglect the running of the coupling and work directly in $d=4$ with arbitrarily small couplings $G$ and $\kappa$.

\begin{figure}[t]
  \hspace{-1cm}  \begin{tabular}{lr}
    \Scale[0.7]
    \input{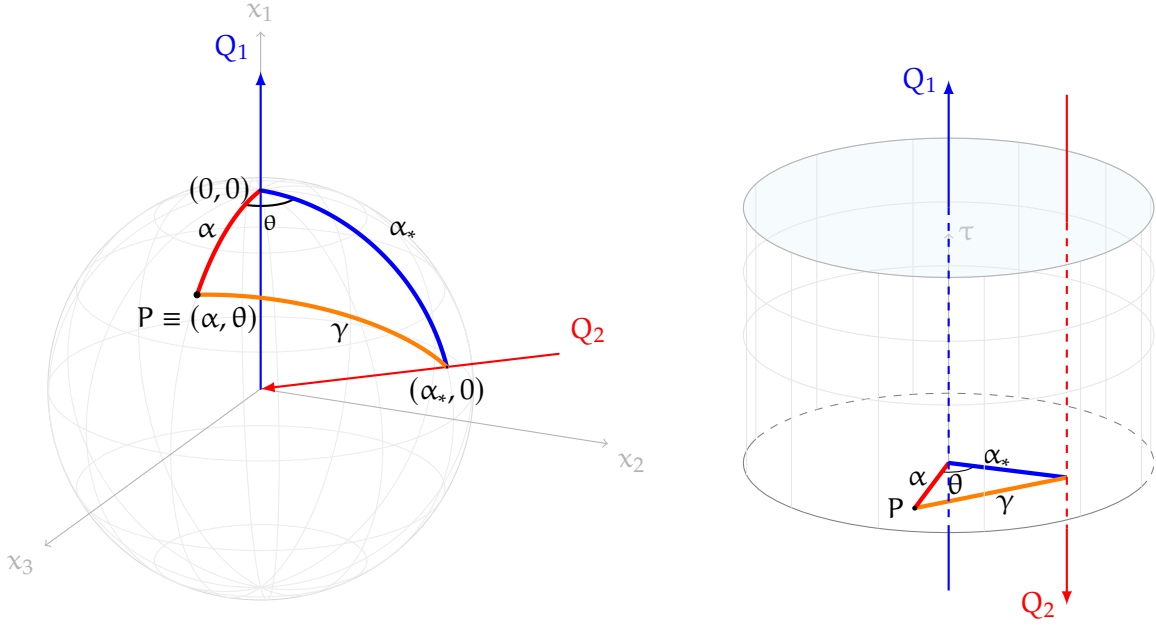} &
                          \Scale[0.9]
                          \tdplotsetmaincoords{70}{110}

\begin{tikzpicture}[tdplot_main_coords, scale=2]

    \def\R{1.5}    %
    \def\H{2}      %
    \def\rInner{1} %
    \def\extra{1}

    \draw[->, gray!60] (0,0,0) -- (0,0,1.8) node[anchor=south, right]{$\tau$};

    \draw[gray] (-90:\R) arc (-90:90:\R);
    \draw[gray, dashed] (90:\R) arc (90:270:\R);

    \coordinate (V1) at (0,0,0);

    \coordinate (V2) at (0,{1.2*\rInner*sin(50)},0);
    \coordinate (V3) at (\rInner,.1,0);

    \draw[ultra thick, blue] (V1) -- (V2);
    \node at ({\rInner / 2 -.1} ,-.1, 0) {\(\alpha\)};

    \draw[ultra thick, orange] (V2) -- (V3);
    \node at ( .7,.7,0) {\(\gamma\)};

    \draw[ultra thick, red] (V3) -- (V1);
    \node at ( -0.2, .3, 0) {\(\alpha_{*}\)};

    \draw (10:.2) arc (10:90:.2);
    \node at ( .4, .2, 0) {\(\theta\)};

    \filldraw (V3) circle (.4pt) node[left] {$P$};

    \foreach \angle in {0,30,...,330}{
        \draw[gray!20, thin] 
            ({\R*cos(\angle)},{\R*sin(\angle)},0) -- ({\R*cos(\angle)},{\R*sin(\angle)},\H);
    }

    \foreach \z in {1,1.5,2}{
        \draw[gray!20, thin] (0,0,\z) circle (\R);
    }

    \fill[cyan!10, fill opacity=0.3] (0,0,\H) circle (\R);
    \draw[gray!60] (0,0,\H) circle (\R);
    \draw[thick, blue, -latex] (0, 0, \H) -- (0, 0, \H + \extra)  node[left] {\(Q_1\)};
    \draw[thick, blue] (0, 0, -\extra) -- (0, 0, -.55) ;
    \draw[thick, blue, dashed] (0, 0, -.55) -- (0, 0, \H) ;
    \draw[thick, red, -latex] (0, {1.2*\rInner*sin(50)}, -.4) -- (0, {1.2*\rInner*sin(50)}, -\extra) node[left] {\(Q_2\)};
    \draw[thick, red, dashed] (0, {1.2*\rInner*sin(50)}, \H) -- (0, {1.2*\rInner*sin(50)}, -.4) ;
    \draw[thick, red] (0, {1.2*\rInner*sin(50)}, \H + \extra) -- (0, {1.2*\rInner*sin(50)}, \H) ;

\end{tikzpicture}    
  \end{tabular}
  \caption{The cusp in \(\setR^4\) (left) and in \(\setR \times S^3\) (right). Because of the invariance under rotations generated by \(\del_\varphi\) we can limit ourselves to the hyperplane \(\varphi=0\). Invariance under \(\del_\tau\) means that we limit ourselves to looking at the two-sphere \(\tau = \text{const}\). In the cylinder picture (right), the two-sphere is represented as a disk of radius \(R \pi\) with the border identified, \emph{i.e.} \(S^2 =D^{2}/ \del D^{2}\). The north pole is the center of the disk and the south pole is its circumference. The two arms of the cusp are mapped to two parallel lines.}
  \label{fig:cusp-sphere-cylinder}
\end{figure}

The cusp preserves dilatations and an \(SO(2)\) rotation.
As usual, it is convenient to pick a coordinate system in which the symmetries are manifest.
In this case, we can use spherical coordinates:
\begin{align}
  \label{eq:spherical-coordinates}
  \begin{dcases}
    x_1 = R e^{\tau/R} \cos(\alpha), \\
    x_2 = R e^{\tau/R} \sin(\alpha) \cos(\theta), \\
    x_3 = R e^{\tau/R} \sin(\alpha) \sin(\theta) \cos(\varphi),\\
    x_4 = R e^{\tau/R} \sin(\alpha) \sin(\theta) \sin(\varphi),\\
  \end{dcases} &\qquad & \begin{dcases}
   \tau \in (-\infty,\infty), \\
    \alpha \in [0,\pi], \\
  \theta \in [0,\pi],\\
  \varphi \in [0,2\pi),
  \end{dcases}
\end{align}
with $R$ a dimensionful parameter.\\
The cusp lives in the plane \(\theta = 0\), with the two arms extended along \(\alpha = 0\) and \(\alpha = \alpha_{*}\).
Dilatations are generated by \(\del_{\tau}\), and the preserved rotations are generated by \(\del_{\phi}\).
We will look for solutions depending only on \(\alpha \) and \(\theta\), assuming that in the regime that we will be considering (small \(G\)), the lowest-energy solution preserves the symmetries.
In these coordinates, the line element is
\begin{equation}
  \dd{s^2} = e^{-2 \tau/R} \left[ \dd{\tau^2} + R^2 \left( \dd{\alpha^2} + \sin^2(\alpha) \left( \dd{\theta^2} + \sin^2(\theta) \dd{\phi^2} \right) \right)\right]  \, ,
\end{equation}
which is manifestly one Weyl transformation away from the cylinder \(\setR_{\tau} \times S^3\). 
After the Weyl transformation, the dilatations are realized as translations in the direction \(\tau\) and the cusp turns into two parallel lines at distance \(\alpha_{*}\) (see Fig.~\ref{fig:cusp-sphere-cylinder}).
It will also be convenient to introduce the (geodesic) angle \(\gamma\) between the \(\alpha_{*}\) arm and a generic line with \((\alpha, \theta)\) fixed.
By the law of cosines,
\begin{equation}
  \label{eq:geodesic-angle}
  \cos(\gamma) = \cos(\alpha) \cos(\alpha_{*}) + \sin(\alpha) \sin(\alpha_{*}) \sin(\theta) \,.
\end{equation}

On the cylinder, cusp operators can be reinterpreted as states in the Hilbert space of two impurities on \(S^{d-1}\). By the state--operator correspondence, \(\Gamma_{q_1 q_2}\) can thus be extracted from the appropriately normalized ground-state energy $E^{\text{cyl}}_{q_1 q_2}$ of the theory on $S^{d-1}$ in the presence of two punctures separated by an angle \(\alpha_*\)~\cite{Correa:2012hh, Cuomo:2024psk}.
The normalization amounts to subtracting the scheme-dependent worldline masses of the individual defects and is fixed by the requirement that a straight conformal line carry zero cusp anomalous dimension, as
\begin{equation}
  -\log\left(\frac{Z_{q_1 q_2}(\alpha_*)}{\sqrt{Z_{q_1q_1}(\pi)Z_{q_2q_2}(\pi)}}\right)=\Gamma_{q_1q_2}(\alpha_*)T \,.
\end{equation}
Here \(T\) is the Euclidean time interval of the cylinder and appears as an overall factor on the right-hand side due to time-translation invariance on the cylinder.
Taking the limit \(T\to\infty\) projects onto the ground state leading to
\begin{equation}
\frac{\Gamma_{q_1 q_2}(\alpha_*)}{R} =  E^{\text{cyl}}_{q_1q_2}(\alpha_*) - \frac{1}{2} E^{\text{cyl}}_{q_1q_1}(\pi) - \frac{1}{2} E^{\text{cyl}}_{q_2q_2}(\pi)   \, .
\end{equation}
In the semiclassical limit, the value of \(\Gamma_{q_1 q_2}\) at the saddle is proportional to \(Q\).
Subleading terms in the \(1/Q\) expansion appear as quantum corrections.
To distinguish these contributions, we introduce the notation
\begin{equation} \label{semicusp}
  \Gamma_{q_1 q_2}(\alpha_{*})=\sum_{i=-1}^\infty \frac{1}{Q^i} \Gamma^{(i)}_{q_1 q_2}(\alpha_{*}) =Q \Gamma^{(-1)}_{q_1 q_2}(\alpha_{*}) + \Gamma^{(0)}_{q_1 q_2}(\alpha_{*}) +  \frac{1}{Q}  \Gamma^{(1)}_{q_1 q_2}(\alpha_{*}) + \dots \,.
\end{equation}

\section{Semiclassical calculation of the  anomalous dimension}
\label{sec:semiclassical-cusp}

\subsection{Equations of motion}
\label{sec:equations-motion}

In the double-scaling limit \(Q \to \infty\), \(\epsilon \to 0\) with \(Q \epsilon\) fixed, the computation of the anomalous cusp dimension becomes semiclassical.
We need to find the saddles of the rescaled action in Eq.~(\ref{eq:cusp-rescaled-action}) and are free to use field redefinitions that might be problematic in a general path integral.

It is convenient to introduce the radial and angular fields \(\rho(x)\) and \(\chi(x)\) and trade \(A\) for the gauge-invariant quantity \(B\):
\begin{align}
  \phi &= \frac{\rho}{\sqrt{2}}e^{i \chi}, & B =i\left( \sqrt{G} A + \dd{\chi}\right),
\end{align}
such that the rescaled action $\bar S$ reads
\begin{multline}
  \label{eq:cusp-rescaled-action-for-B}
 \bar S  =  \int \dd{x} \Big(- \frac{1}{4G}   \tilde F_{\mu\nu}   \tilde F^{\mu\nu} +\frac12 \left(\del \rho\right)^2-\frac12 \rho^2 B^\mu B_\mu+ \frac{\kappa}{4} \rho^4
  + \left( q_2 - q_1 \right) \log( \rho )\pqty*{ \delta(x) + \delta(x - x_{\infty}) } \Big)\\- q_1 \int_{\gamma_1} B + q_2  \int_{\gamma_2} B \, ,
\end{multline}
where we set $\phi_{a\ge 2}=0$ and
\begin{equation}
  \tilde F = \dd{B} =i \sqrt{G} F\,.
\end{equation}
The \ac{eom} in covariant form read
\begin{align}
  \label{eq:EOM1}
  \frac{1}{\sqrt{g}}  \partial_\mu \left(\sqrt{g}\tilde{F}^{\mu \nu}  \right)&= G \left(\rho ^2 B^{\nu } +q_1 J_{\text{MS}}^{\nu }(\gamma_1) - q_2 J_{\text{MS}}^{\nu} (\gamma_2)\right)\,,  \\
  \label{eq:EOM2}
  \frac{1}{\sqrt{g}}  \partial_\mu \left(\sqrt{g}g^{\mu \nu }\partial_\nu \rho \right)&=  \left(-B^{\mu } B_{\mu } +\kappa  \rho ^2\right) \rho +  \frac{(q_2-q_1)}{\rho }\pqty{\delta^d\left(x\right) + \delta^d\left(x - x_{\infty}\right)} \,,
\end{align}
where \(J_{\text{MS}}\) are the \ac{ms} currents for the two arms of the cusp as defined in Eq.~\eqref{eq:JMS}.
With these sources, the Gauss law takes the form
\begin{equation}
  \label{eq:Gauss-law}
  \int  \sqrt{g}\left(\rho ^2 B^{\mu }(x) +  q_1 J_{\text{MS}}^{\mu}(x \mid \gamma_1) - q_2 J_{\text{MS}}^{\mu }(x \mid \gamma_2) \right)  \dd{x} = 0 \, .
\end{equation}

The presence of the cusp leads naturally to the issue of the choice of appropriate boundary conditions for the scalar field.
In fact, there are essentially just two choices, related to the boundary terms that can be added to the action.
We choose the boundary condition that imposes regular behavior for \(\rho\) at the cusp, assuming that, at least in the perturbative small-\(G\) regime, it will lead to a lower energy configuration.
We will return to this issue with a more detailed discussion in Section~\ref{sec:boundary-terms}.

Having written the \ac{eom} in covariant form we are free to use the spherical coordinate system of Eq.~(\ref{eq:spherical-coordinates}) in which the symmetries of the problem are manifest.
The insertions of the scalar field sit at \(\tau \to \pm \infty\), so both delta functions in the action of Eq.~\eqref{eq:cusp-rescaled-action} can be traded for boundary conditions.
As already observed above, the arms of the cusp are parallel to the \(\tau\) axis and the corresponding current \(J_{\text{MS}}\) has only one non-vanishing component: for a line at fixed solid angle \(\Omega_{*}\) , it is simply
\begin{equation}
  J_{\text{MS}} = J_{\text{MS}}^{\tau} \del_{\tau}{} = - \frac{\delta(\Omega - \Omega_{*})}{R^3} \del_{\tau}{} \, .
\end{equation}
Having assumed that there is no spontaneous symmetry breaking, the Maxwell equation~\eqref{eq:EOM1} implies that also the \(B\) field is oriented in the \(\tau\) direction: \(B = B^{\tau} \del_{\tau}{} \equiv B \del_{\tau}{}\).

The expressions simplify if we rescale the field \(\rho\) as
\begin{equation}
  \rho = e^{-\tau/R} \rho_c \, ,  
\end{equation}
which corresponds to the Weyl rescaling that turns \(\setR^4\) into \( \setR \times S^3\).
Like this, the dilatation invariance is realized if we demand both \(B\) and \(\rho_c\) to be functions only of the angles \(\alpha\) and \(\theta\).

With these simplifications, the \ac{eom} take the form
\begin{gather} 
  \Laplacian_{S^3} \rho_c(\alpha, \theta) = \left(1-(R B(\alpha, \theta))^2+\kappa  (R \rho_c(\alpha, \theta)) ^2\right) \rho_c(\alpha, \theta)  , \\
  \Laplacian_{S^3}B(\alpha, \theta) = G \left((R \rho_c(\alpha, \theta))^2 B(\alpha, \theta)  - \frac{q_1 \delta(\alpha)- q_2 \delta (\alpha-\alpha_* ) \delta(\theta)}{2 \pi R \sin ^2\alpha \sin\theta }\right),  \\
  \int_0^\pi \sin(\theta) \dd{\theta}  \int_0^\pi \sin^2(\alpha) \dd{\alpha} \rho_c^2(\alpha, \theta) B(\alpha, \theta)  =\frac{q_1 - q_2 }{2 \pi R^3} \, ,
\end{gather}
where \(\Laplacian_{S^3}{}\) is the Laplacian on the three-sphere
\begin{equation}
   \Laplacian_{S^3}{ } =  \tfrac{1}{\sin^2 \alpha}\del_\alpha\left(\sin^2 \alpha \del_\alpha{} \right) + \tfrac{1}{\sin^2 \alpha \sin \theta}\del_\theta\left(\sin \theta \del_\theta {} \right) + \tfrac{1}{\sin^2 \alpha \sin^2\theta}\del_\phi^2 {} \, .
\end{equation}
The saddle-point action reduces to
\begin{multline}
    E_{q_1 q_2}^{(-1)}(\alpha_*) \simeq \frac{\bar S_{\saddle}}{T} =  \pi R  \int_0^{\pi }  \sin(\theta) \dd{\theta} \int_0^{\pi } \sin ^2(\alpha ) \dd{\alpha} \Bigg[\frac{1}{G}B\Laplacian_{S^3}B-\rho_c\Laplacian_{S^3}\rho_c \\
  +\left(1-(R B)^2 +\frac{\kappa}{2}(R\rho_c)^2\right)\rho_c ^2+\frac{q_1 \delta (\alpha ) - q_2 \delta (\theta ) \delta (\alpha -\alpha_*)}{\pi R \sin ^2(\alpha ) \sin (\theta )}B \Bigg] \,.
\end{multline}
The current insertions on the right-hand side of the Maxwell equation impose the following boundary conditions on the gauge field near the arms of the cusp:
\begin{align}
  \frac{\del B}{\del \alpha} \xrightarrow[\alpha \to 0]{} -\frac{G q_1}{4 \pi R \alpha^2} \,, &&  \frac{\del B}{\del \gamma} \xrightarrow[\gamma \to 0]{} \frac{G q_2}{4 \pi R \gamma^2} \, ,
\end{align}
where \(\gamma\) is the geodesic angle in Eq.~(\ref{eq:geodesic-angle}).

\subsection{Boundary terms and boundary conditions}
\label{sec:boundary-terms}

To solve the \ac{eom} we need to specify boundary conditions for the scalar field \(\rho_c\) at the cusp.
Different choices of boundary conditions will then correspond to different localized boundary terms in the action.

Expanding the \ac{eom} close to the cusp for $G<2\pi$, one finds the two possible behaviors
\begin{equation}
  \begin{cases}
    B =\frac{q_1 G}{4\pi R\alpha}, \quad \rho_c \sim \rho_{\pm}(\kappa, G) \alpha^{-(1\pm\nu_1)/2} &\text{for \(\alpha \sim 0\)},\\
    B =-\frac{q_2 G}{4\pi R\gamma}, \quad \rho_c \sim \rho_{\pm}(\kappa, G) \gamma^{-(1\pm\nu_2)/2} &\text{for \(\alpha \sim \alpha_{*}\)},
  \end{cases}
\end{equation}
where
\begin{align}
	\nu_{1}&=\sqrt{1 - \frac{(q_1G)^2}{4\pi^2}} \,, &  \nu_{2}&=\sqrt{1 - \frac{(q_2G)^2}{4\pi^2}}\,,  
\end{align}
and $\rho_\pm(\kappa,G)$ are so-far-undetermined angle-independent coefficients, analytic in the couplings.
Given the obvious symmetry, we can concentrate on the $\alpha=\alpha_{*}$ arm.
Expanding in powers of $G$ and noting that
\begin{align}
	 \gamma^{-(1+\nu_2)/2} &\sim  \frac{1}{\gamma} + \frac{(q_2G)^2}{16\pi^2} \frac{\log \gamma}{\gamma}+ \dots \,, &  \gamma^{-(1-\nu_2)/2} &\sim  1 - \frac{(q_2G)^2}{16\pi^2} {\log \gamma}+ \dots \,,
\end{align}
one can see that the $\rho_-$ branch is smoothly connected in the limit $G\to0$ to the homogeneous solution where the scalar has a constant profile.

To find the corresponding boundary terms in the action, we start by observing that if we do not neglect total derivatives in the variation of the bulk action, and impose a \ac{uv} cutoff $\gamma_0$ away from the cusp location, then the variation includes a boundary-localised term 
\begin{equation}
	\eval*{\frac{\delta S_{\text{bulk}}}{T}}_{\gamma=\gamma_0} = \eval*{4\pi R \sin^2\gamma \del_\gamma \rho_c \delta\rho_c }_{\gamma=\gamma_0} \,.
\end{equation}
We can compensate it with a term localized on the second arm of the line,
\begin{equation}
	S_{\text{bdry}} = \eval*{\pi R F\int \gamma \rho_c^2 \dd{\tau}}_{\gamma=\gamma_0},
\end{equation}
where \(F\) is a new, running, coupling~\cite{Gubser:2002vv,Klebanov:1999tb,Aharony:2022ntz,Aharony:2023amq}.
The total variation is then proportional to
\begin{equation}
    \eval*{\delta S_{\text{bulk}}}_{\gamma=\gamma_0} + \delta S_{\text{bdry}} \propto  \rho_+ \gamma_0^{-\nu_2} \pqty*{F -  \pqty*{1 + \nu_2}} +  \rho_- \pqty*{F -  \pqty*{1 - \nu_2}} \, ,
\end{equation}
and vanishes if
\begin{equation}
  \frac{\rho_+}{\rho_-} = - \frac{F -1 - \nu_2}{F -1 +\nu_2} \gamma_0^{-\nu_2} \, .
\end{equation}
Imposing the \ac{rhs} to be cutoff-independent leads to the beta function for \(F\):
\begin{equation}
  \beta_F = \gamma_0 \odv{F}{\gamma_0} = \frac{(q_2 G)^2}{8 \pi} - F + \frac{F^2}{2} \, .
\end{equation}
The flow joins two fixed points:
\begin{align}
   F_{\ac{ir}} &=1 - \nu_2,  &  F_{\ac{uv}} &= 1 + \nu_2,
\end{align}
which correspond, respectively, to the vanishing of \(\rho_+\) and \(\rho_-\).
In either case, the boundary term evaluated at the saddle does not contribute to the total energy.

In the following, we will concentrate on the \(\rho_-\) branch that is, as observed above, continuously connected to a constant solution for \(G =0\). 

The analysis changes radically if \(G > 2 \pi\).
In this case, the behavior at the cusp becomes
\begin{equation}
  \rho_c \underset{\gamma \to 0}{\sim} \frac{1}{\gamma^{1/2}R} \cos\left(\frac{\abs{\nu_2}}{2}  \log\frac{\gamma}{\gamma_0}\right)  \,.
\end{equation}
Correspondingly, the two fixed points in the \ac{rg} flow of the coupling \(F\) merge at \(G = 2\pi\) and disappear in the complex plane.
Everything points towards an instability of the classical solution, with the formation of a new screening condensate with dimensional transmutation taking place, in analogy with the discussion in~\cite{Aharony:2022ntz,Aharony:2023amq}.
We therefore expect a new strongly-coupled phase in which more symmetries might be broken.

\subsection{Leading order in \(G\)}
\label{sec:leading-order-g}

In spite of the double-scaling limit we are still dealing with a set of non-linear \acp{pde}.
To solve them in a useful regime, we choose to concentrate on the limit \(G \to 0\) with \(\kappa\) fixed.
This regime overlaps with and extends the standard perturbative regime.
It allows us on the one hand to validate our approach, and on the other, to extend known results in the literature.

To find a perturbative solution in \(G\), we expand the fields as
\begin{align}
  \braket{B (\alpha, \theta) } &\equiv \mu(\alpha,\theta) = \mu_0 + G \mu_1(\alpha,\theta) + G^2 \mu_2 (\alpha,\theta)+ \dots \,, \\
  \braket{\rho_c(\alpha, \theta)} &\equiv r(\alpha,\theta) = r_0 + G r_1(\alpha,\theta) + G^2 r_2(\alpha,\theta)  + \dots,
\end{align}%
where we have anticipated that for $G=0$ the lowest-energy solution is spatially homogeneous as the line disappears.
The energy on the cylinder will have a similar expansion in $G$ at each order in \(Q\), for which we introduce a second index
\begin{equation}
  \Gamma^{(i)}_{q_1q_2}(\alpha_*) = \sum_{j=0}^\infty G^j\Gamma_{q_1q_2}^{(i, j)}(\alpha_*)=\Gamma_{q_1q_2}^{(i, 0)}(\alpha_*)+ G \Gamma_{q_1q_2}^{(i, 1)} (\alpha_*) + G^2 \Gamma_{q_1q_2}^{(i, 2)} (\alpha_*)+\dots
\end{equation}

At leading order in $G$ and \(Q\), the computation parallels the one for the scaling dimension of the operator $\phi^{Q_1-Q_2}$ in the (ungauged) critical $O(2N)$ $\phi^4$ theory~\cite{Badel:2019oxl,Antipin:2020abu}, in which $\mu_0$ has the interpretation of a chemical potential.
In this case, the \ac{eom} reduce to
\begin{align}
  \label{eq:LOparam}
  2 \pi ^2 \mu_0  \left(\mu_0 ^2-1\right)&= \kappa (q_1-q_2) \,, & r_0 &=\frac{1}{\pi R}\sqrt{\frac{q_1-q_2}{2 R \mu_0}},
\end{align}
and the corresponding energy is given by
\begin{equation} \label{primo}
  \Gamma_{q_1 q_2}^{(-1,0)} = \frac{q_1-q_2}{4}\left( 3 R \mu_0+\frac{1}{R \mu_0 }\right) \,.
\end{equation}
The cubic equation for $\mu_0$ can be solved to give an exact expression for \(\Gamma_{q_1 q_2}^{(-1, 0)}\) that depends only on the difference \(q_1 - q_2\), and $\kappa$ (see Appendix~\ref{sec:explicit} for an explicit expression).
For later convenience we report the exact expression for \(\mu_0\):
\begin{equation}
  \label{eq:mu0}
  \mu_0 = \frac{2 ( 6 \pi^4)^{1/3} +\left(\sqrt{81 \kappa ^2 (q_1 - q_2)^2-48 \pi ^4}+9 \kappa  (q_1 - q_2)\right)^{2/3}}{(6 \pi )^{2/3} \left(\sqrt{81 \kappa ^2 (q_1 - q_2)^2-48 \pi ^4}+9 \kappa  (q_1 - q_2)\right)^{1/3}} \, .
\end{equation}

The behavior becomes more transparent if we take the limits of large and small double-scaling parameter \(\kappa\):
\begin{equation}
  \label{eq:Gamma-10-expansions}
   \Gamma_{q_1 q_2}^{(-1,0)} \sim
  \begin{cases}
    \pqty{q_1 - q_2} + \frac{\pqty{q_1 - q_2}^2 \kappa }{8 \pi ^2} - \frac{\pqty{q_1 - q_2}^3 \kappa ^2}{32 \pi ^4} + \frac{\pqty{q_1 - q_2}^4 \kappa ^3}{64 \pi ^6} + \dots & \text{for \(\kappa \ll 1\)} \\
    \frac{3 \pqty{q_1 - q_2}^{4/3}\kappa^{1/3}}{2^{7/3} \pi ^{2/3}} + \left(\frac{\pi }{2}\right)^{2/3} \pqty{q_1 - q_2}^{2/3}\kappa^{-1/3} - \frac{\pi ^2}{6 \kappa } + \frac{\pi ^{10/3} \kappa^{-5/3}}{ 2^{1/3} 9 \pqty{q_1 - q_2}^{2/3}} + \dots & \text{for \(\kappa \gg 1\)} \,.
  \end{cases}
\end{equation}
As expected from general considerations in the large charge expansion, \(\Gamma_{q_1 q_2}^{(-1,0)}\) interpolates between \(q_1 - q_2\) at small \(\kappa\) and \(\pqty{q_1 - q_2}^{4/3}\) at large \(\kappa\).
In the special case of \(q_1 = q_2\), we find that the leading-order term in the expansion vanishes: \(\Gamma^{(-1,0)}_{q q}  = 0 \).

\subsection{\Acl{nlo}}

The first non-trivial term in the small-\(G\) expansion of the energy is given by the integral
\begin{equation}
    E_{q_1q_2}^{(-1,1)}(\alpha_*) = \frac{1}{2 \pi} \int _0^{\pi} \sin(\theta) \dd{\theta} \int _0^{\pi } \sin ^2(\alpha ) \dd{\alpha}  \left(q_2 - q_1 + \frac{q_{1 }\delta(\alpha) - q_2 \delta (\alpha -\alpha_*) \delta (\theta) }{\sin ^2(\alpha ) \sin (\theta )}\right) \mu_1(\alpha ,\theta ).
\end{equation}
The on-shell action depends only on $\mu_1$ which is determined by solving the equation
\begin{equation}
  \Laplacian_{S^3}\mu_1(\alpha,\theta) = \frac{q_1-q_2}{2 \pi^2 R} - \frac{q_1 \delta(\alpha)}{4 \pi R \sin^2 \alpha}+ \frac{q_2 \delta(\alpha-\alpha_*)\delta(\theta)}{2 \pi R \sin^2 \alpha \sin\theta} \,.
\end{equation}
As discussed in Section~\ref{sec:boundary-terms}, we choose boundary terms so that the solution smoothly connects to a homogeneous charge distribution in the limit \(G \to 0\).
For \(\mu_1\), this boundary condition leads to the expression
\begin{equation}
  \label{eq:mu1}
  \mu_1(\alpha, \gamma) =-\frac{q_1-q_2}{8\pi^2R}+\frac{q_1 \pqty{\pi -\alpha} \cot \alpha - q_2 \pqty{\pi -\gamma}  \cot (\gamma )}{4 \pi ^2 R} \, .
\end{equation}
It is convenient to use the geodesic angle \(\gamma \) of Eq.~\eqref{eq:geodesic-angle} since, due to the geometry of the problem and the superposition principle for the linearized equations, the perturbative solutions of the \ac{eom} always take the form $q_1 f(\alpha) - q_2 f(\gamma)$.

\begin{figure}
  \centering
  \begin{tabular}{llr}
    \includegraphics[width=.42\textwidth]{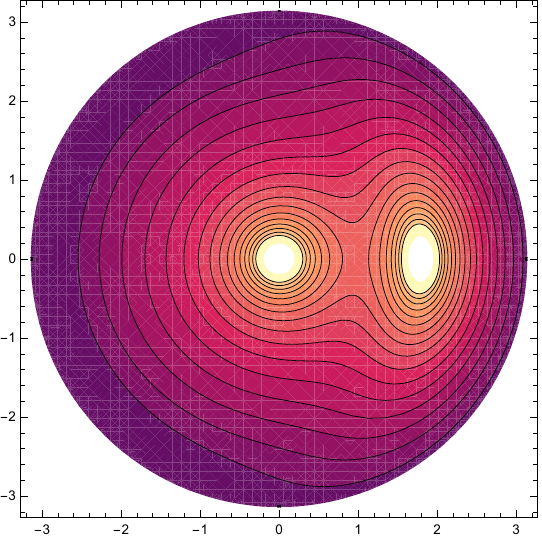} &&                                                                                        \includegraphics[width=.42\textwidth]{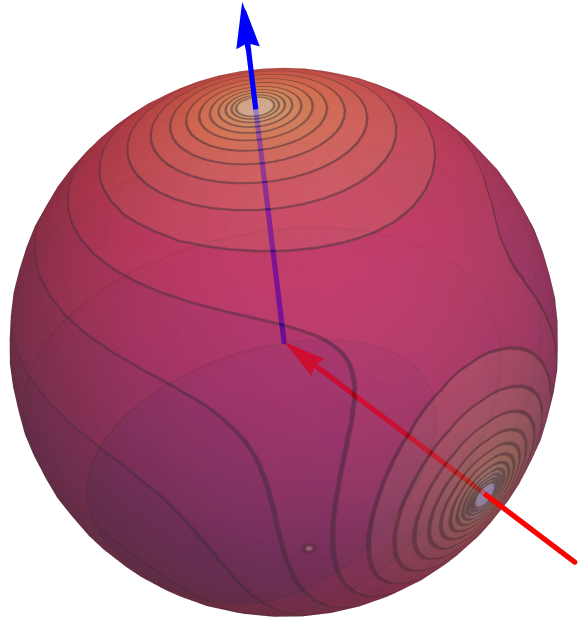}
  \end{tabular}
  \caption{Energy density at first order in \(G\) in units of \(Q\) for \(q_2/q_1 = 2/3\). Here we show the density on a two-sphere at fixed values of \(r\) and \(\varphi\) in stereographic (left) and perspective (right) projection. The density is strongly peaked along the arms of the cusp.}
  \label{fig:energy-density-G}
\end{figure}

Unsurprisingly, the energy density is peaked around the cusp and decreases very rapidly away from it (see Figure~\ref{fig:energy-density-G}).

\begin{figure}
  \centering
  \includegraphics[width=.4\textwidth]{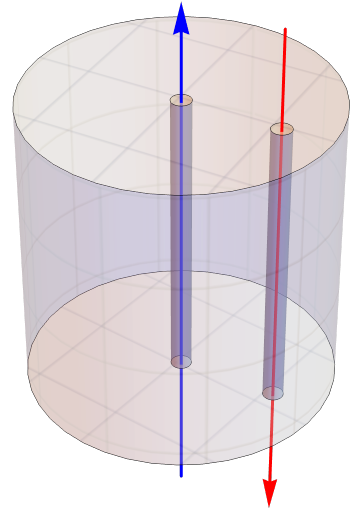}
  \caption{The energy density integral is regulated by removing two cylinders of radius \(\hat \delta \to 0\) around the arms of the cusp.}
  \label{fig:regulated-cylinder}
\end{figure}

As it is, the integral for the energy has an \ac{uv} divergence originating from the region close to the arms of the cusp.
To regulate it we need to introduce both a \ac{uv} regulator \(\hat \delta\) which in the cylinder frame corresponds to removing two infinite cylinders of radius \(\hat \delta\) around the Wilson lines (Figure~\ref{fig:regulated-cylinder}), and a mollifier \(\eta_{\delta}\) for the \(\delta\) function.
The simplest option is
\begin{equation}
  \eta_{\delta}(\alpha) =
  \begin{cases}
    1/\delta  & \text{if \(0 < \alpha < \delta\)} \\
    0 & \text{otherwise}\,, 
  \end{cases}
\end{equation}
with \(\hat \delta < \delta\).
Then, around \(\alpha = 0\) we trade the divergent part of the integral for
\begin{equation}
  \frac{q_1^2}{8 \pi} \int_0^{\pi} \dd{\alpha} \frac{1}{\alpha} \delta(\alpha) \mapsto \lim_{\delta \to 0} \frac{q_1^2}{8 \pi} \int_{\hat \delta}^{\pi} \dd{\alpha} \frac{1}{\alpha} \eta_{\delta}(\alpha) = \frac{q_1^2}{8 \pi} \log(\delta/\hat \delta) \frac{1}{\delta}\,,
\end{equation}
which shows manifestly that the integral diverges as \(1/\delta\) in the limit \(\delta \to 0\), \(\hat \delta \to 0\) with \(\delta/ \hat \delta\) fixed.
The same reasoning applies to the divergence in \(\alpha = \alpha_{*}\), which is proportional to \(q_2^2\) with the same numerical coefficient.
The final result is that the regulated energy has the form
\begin{equation}
  R E_{q_1 q_2}^{(-1,1)}(\alpha_{*} \mid \delta, \hat \delta ) = \pqty*{q_1^2 + q_2^2} \frac{\log(\delta/ \hat \delta)}{8 \pi  \delta } - \frac{1}{16 \pi^2} \left( 3\left(q_1 - q_2 \right)^2 + 4  \left( 1 + \left( \pi - \alpha_{*} \right) \cot(\alpha_{*}) \right) q_1 q_2 \right) \, ,
\end{equation}
and the cusp anomalous dimension is regular, since we subtract the contribution of two infinite lines with charge \(q_1\) and \(q_2\):
\begin{equation}
  \Gamma_{q_1 q_2}^{(-1,1)}(\alpha_{*}) = R \lim_{\delta \to 0} \lrp*{E_{q_1 q_2}^{(-1,1)}(\alpha_{*} \mid \delta, \hat \delta) - \frac{1}{2} E_{q_1 q_1}^{(-1,1)}(\pi \mid \delta, \hat \delta) - \frac{1}{2} E_{q_2 q_2}^{(-1,1)}(\pi \mid \delta, \hat \delta)} .  
\end{equation}
All in all, we find the \(\order{G}\) correction to the cusp dimension to be
\begin{equation} \label{eq:NLOcusp}
    \Gamma_{q_1 q_2}^{(-1,1)} = - \frac{1}{16 \pi^2} \left( 3\left(q_1 - q_2 \right)^2 + 4  \left( 1 + \left( \pi - \alpha_{*} \right) \cot(\alpha_{*}) \right) q_1 q_2 \right) \, ,
  \end{equation}
which is manifestly invariant under the symmetries \((q_1, q_2) \to ( -q_2, -q_1)\), and \(\alpha_* \to 2 \pi - \alpha_*\).

Despite the simple form of this result, it represents an all-order statement from the point of view of conventional perturbation theory: Eq.~\eqref{eq:NLOcusp} implies that in the diagrammatic expansion of $\Gamma_{q_1 q_2}$ all the terms scaling as $Q G \kappa^k = Q^{2+k} e^{2}\lambda^k$ are zero for any $k>0$, as will be discussed later in Section~\ref{sec:pert-epsil-expans}.

\subsection{\Acl{nnlo}}

\begin{figure}
  \centering
  \begin{tabular}{lr}
    \begin{tikzpicture}
      \node[anchor=south west, inner sep=0] (image) at (0,0)
      {\includegraphics[width=0.4\textwidth]{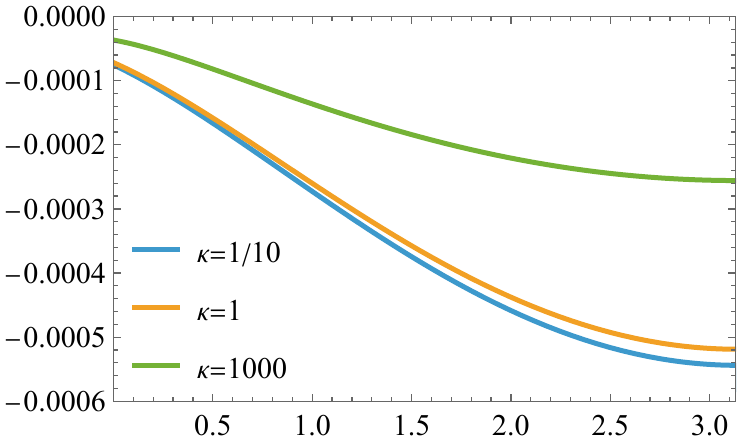}};
      \begin{scope}[x={(image.south east)}, y={(image.north west)}]
        \node[] at (-.075, 0.5) {\footnotesize \(\Gamma^{(-1,2)}_{q_1 q_2}\)};
        \node[] at (.5, -.1) {\footnotesize \(\alpha_{*}\)};
      \end{scope}
    \end{tikzpicture}
     &
    \begin{tikzpicture}
      \node[anchor=south west, inner sep=0] (image) at (0,0)
      {\includegraphics[width=0.41\textwidth]{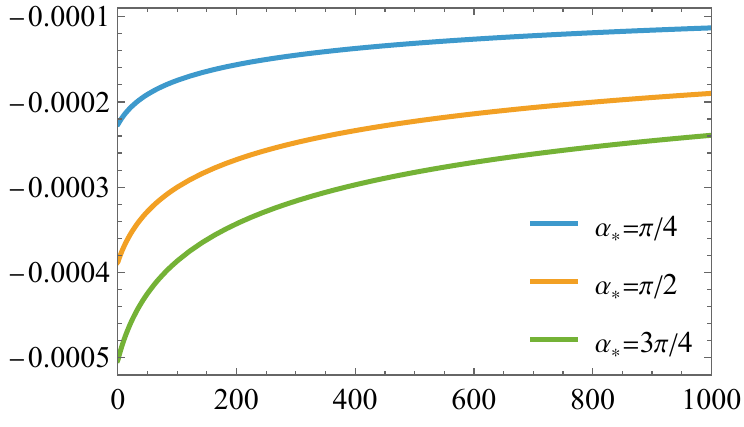}};
      \begin{scope}[x={(image.south east)}, y={(image.north west)}]
        \node[] at (-.075, 0.5) {\footnotesize \(\Gamma^{(-1,2)}_{q_1 q_2}\)};
        \node[] at (.5, -.1) {\footnotesize \(\kappa\)};
      \end{scope}
    \end{tikzpicture}
  \end{tabular}
  \caption{NNLO contribution to the cusp anomalous dimension $\Gamma_{q_1q_2}^{(-1,2)}(\alpha_*)$ as a function of the cusp angle (\emph{left}) and $\kappa$ (\emph{right}). We considered $q_1=1$ and $q_2=1/2$. Nondegenerate different values of the charges lead to a qualitatively similar behavior.}\label{fig:NNLOcusp} 
\end{figure}
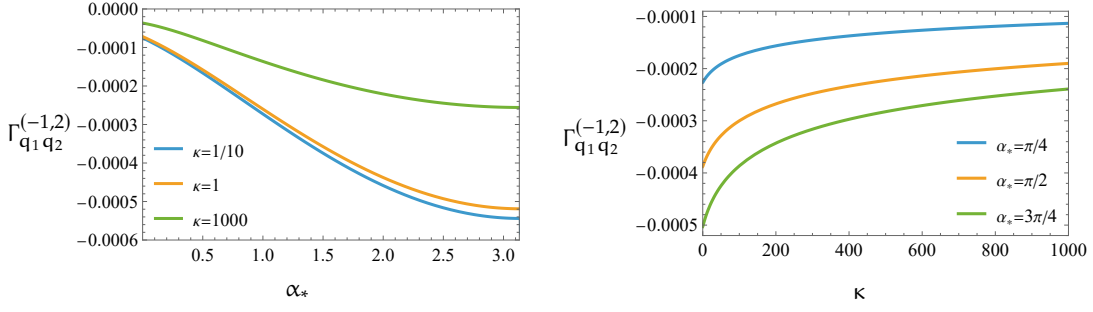

Using the \ac{eom} to simplify the on-shell action, we arrive at the following integral expression for the \ac{nnlo} contribution to the energy:
\begin{multline}
  E_{q_1q_2}^{(-1,2)}(\alpha_*) = -  \int _0^{\pi } \sin (\theta ) \dd{\theta}\int _0^{\pi }\sin^2(\alpha ) \dd{\alpha} \\
  \left(\frac{\pqty*{q_1 - q_2} \mu_1(\alpha ,\theta )}{2 \pi  \mu_0}+R \sqrt{2 \pqty*{q_1 - q_2} R\mu_0 } r_1(\alpha ,\theta )\right)\mu_1(\alpha ,\theta ),
\end{multline}
where \(\mu_0 = \mu_0(q_1 - q_2)\) is the expression that we have reported in Eq.~(\ref{eq:mu0}), and $r_1$ solves
\begin{equation}
 \Laplacian_{S_3}r_1(\alpha,\theta) =  2 \left((R\mu_0)^2-1\right) r_1(\alpha,\theta) -\frac{\sqrt{ 2 \pqty*{q_1 - q_2} R \mu_0 }}{\pi } \mu_1(\alpha,\theta) \,.
\end{equation}
The regular Green's function solution for $r_1$ reads
\begin{multline}
  r_1(\alpha, \gamma) =  \frac{\sqrt{\pqty*{q_1 - q_2} R\mu_0 }}{4 \sqrt{2} \pi ^3 R \left((R\mu_0)^2-1\right)}\Bigg[
  \frac{\pqty*{3-(R\mu_0)^2} \pqty*{q_1 - q_2}}{2\left( (R\mu_0)^2-1\right)}
  + q_1 \pqty*{\pi -\alpha} \cot (\alpha ) - q_2 \pqty*{\pi - \gamma} \cot (\gamma ) \\
 - \tfrac{\pi}{\sinh*(\pi  \sqrt{2(R \mu_0)^2-3})}
  \Bigg(
  q_1 \tfrac{\sinh*( \pqty*{\pi - \alpha}  \sqrt{2 (R \mu_0)^2-3})}{\sin(\alpha)}
  - q_2 \tfrac{\sinh*( \pqty*{\pi - \gamma}  \sqrt{2 (R \mu_0)^2-3})}{\sin(\gamma)}
  \Bigg)
  \Bigg] \,.
\end{multline}
The integral over the energy density can be easily evaluated by exploiting the symmetry under the exchange of the two arms of the cusp line.
At this order, the solution is lengthy and not a particularly transparent function expressed in terms of $\mu_0$. While we have an explicit expression for the result (Appendix~\ref{sec:explicit}), since it depends on the coupling \(\kappa\), we find it more illuminating to give the small and large \(\kappa\) expansions.
For \(\kappa \ll 1\): 
\begin{multline}
  \label{eq:Gamma-12-expansion-small}
    \Gamma_{q_1 q_2}^{(-1,2)}(\alpha_{*}) = \frac{\left(3-8 \pi ^2\right) }{384 \pi ^4} (q_1-q_2)^3 +  \frac{ \alpha_{*} (\alpha_{*}-2 \pi ) (\alpha_{*}-\pi ) \cot (\alpha_{*})-2 \pi ^2 }{48 \pi ^4}  \pqty{q_1 - q_2} q_1 q_2  \\
+
\frac{45-15 \pi ^2+2 \pi ^4 }{2880 \pi ^6} \pqty{q_1 - q_2}^4 \kappa \\
  + \tfrac{ \alpha_{*}^4-4 \pi  \alpha_{*}^3+4 \left(\pi ^2-3\right) \alpha_{*}^2+24 \pi  \alpha_{*}-12 \pi ^2 + 6 (\alpha_{*}-2 \pi ) (\alpha_{*}-\pi ) \alpha_{*} \cot (\alpha_{*})}{384 \pi ^6} \pqty{q_1 - q_2}^2 q_1 q_2  \kappa \\
  + \order{\kappa^2} \,, 
\end{multline}
and for \(\kappa \gg 1\): 
\begin{multline}
  \label{eq:Gamma-12-expansion-big}
  \Gamma_{q_1 q_2}^{(-1,2)}(\alpha_{*}) = -\pqty*{\tfrac{ 3+4 \pi ^2}{64 \times 2^{2/3} \pi ^{10/3}} \pqty{q_1 - q_2}^2 + \tfrac{3(2 \pi -\alpha_{*}) \alpha_{*} }{16 \times 2^{2/3} \pi ^{10/3}} q_1 q_2 } \frac{\pqty{q_1 - q_2}^{2/3}}{\kappa^{1/3}}\left(1+ \order{\kappa^{-1/3}}\right) \\
  +\frac{1}{4 \pi  \kappa  \sin (\alpha_*)}\exp*[-\sqrt{2}\left(\frac{ \kappa \pqty{q_1 - q_2}}{2\pi ^2}\right)^{1/3}\alpha_* ] \left(1+ \order{\kappa^{-2/3}}\right) + \dots ,
\end{multline}
where the dots denote subleading exponentially suppressed contributions.
At this order, the dependence of $\Gamma_{q_1 q_2}$ on the cusp angle $\alpha_*$ is nontrivially determined by the quartic coupling.
As shown in Fig.~\ref{fig:NNLOcusp}, as $\kappa$ increases, the dependence on $\alpha_*$ becomes progressively milder and $\Gamma_{q_1 q_2}^{(-1,2)}$ goes to zero as $\kappa^{-1/3}$.

\bigskip

\begin{figure}
  \centering
  \begin{tikzpicture}
    \node[anchor=south west, inner sep=0] (image) at (0,0)
    {\includegraphics[width=0.6\textwidth]{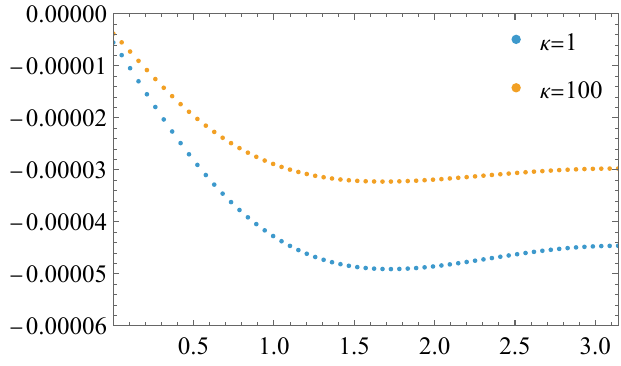}};
    \begin{scope}[x={(image.south east)}, y={(image.north west)}]
      \node[] at (-.05, 0.5) { \(\Gamma^{(-1,3)}_{q_1 q_2}\)};
      \node[] at (.5, -.05) { \(\alpha_{*}\)};
    \end{scope}
    \end{tikzpicture}
  \caption{NNNLO contribution to the cusp anomalous dimension $\Gamma_{q_1q_2}^{(-1,3)}(\alpha_*)$ as a function of the cusp angle $\alpha_*$ for $\kappa=1$ and $\kappa=100$. The considered values of the charges are $q_1=1$ and $q_2=1/2$.}\label{fig:NNNLOcusp} 
\end{figure}

We conclude the section by providing an expression for $\Gamma^{(-1,3)}_{q_1 q_2}$ which can be easily evaluated numerically. We have 
\begin{multline}
   \Gamma^{(-1,3)}_{q_1 q_2} = \int _0^{\pi } \sin \theta \dd{\theta} \int _0^{\pi }\sin ^2\alpha \dd{\alpha} \Bigg[ 2 \pi ^2 \left((R\mu_0)^2-1\right)  \sqrt{\frac{2 R \mu_0}{q_1-q_2}} (R r_1)^3-\frac{(q_1-q_2) R\mu_1 \mu_2}{2 \pi  \mu_0}\\
  - \sqrt{2 \pqty{q_1-q_2} R \mu_0 }\left((R\mu_1)^2+R \mu_2\right) Rr_1 -2 \pi R^4 \mu_0 \mu_1  r_1^2\Bigg]  ,
\end{multline}
where \(\mu_1 \) is a function of $q_1$, $q_2$, and \(\kappa\) as in Eq.~(\ref{eq:mu1}) and%
\footnote{In solving the equation of motion for $\mu_2$ we are left with one unfixed integration constant, which can be fixed using charge conservation if the solution $r_2$ is known. However, since $\Gamma_{q_1q_2}^{(-1,3)}$ is independent of this parameter, we here fixed it to a convienient value.}
\begin{multline}
  \mu_2(\alpha, \gamma) = \tfrac{q_1-q_2}{32 \pi ^4 R^2 \mu_0  \left((R\mu_0)^2-1\right)^2}\Bigg[ q_1 \Bigg( 4 \pqty{\pi -\alpha}  (R\mu_0) ^2 \cot (\alpha ) + \pqty{2 \pi -\alpha } \alpha  \pqty{3 (R \mu_0) ^4-4( R \mu_0 )^2 + 1 }\\
  - \tfrac{4 \pi (R \mu_0 )^2 \sinh*((\pi -\alpha ) \sqrt{2( R \mu_0) ^2-3})}{\sin(\alpha ) \sinh*(\pi  \sqrt{2 (R \mu_0) ^2-3} ) } \Bigg) \\
-  q_2 \bigg( 4 \pqty{\pi -\gamma}  \mu_0 ^2 \cot (\gamma ) + \pqty{2 \pi -\gamma } \gamma  \pqty{3 (R \mu_0) ^4-4 (R\mu_0 )^2 + 1 }
\\
- \tfrac{4 \pi  (R\mu_0 )^2 \sinh*((\pi -\gamma ) \sqrt{2 (R\mu_0) ^2-3})}{\sin(\gamma ) \sinh*(\pi  \sqrt{2 (R\mu_0) ^2-3} ) } \bigg)
 \Bigg].
\end{multline}
In Fig.~\ref{fig:NNNLOcusp} we illustrate the $\alpha_*$ dependence of $\Gamma^{(-1,3)}_{q_1 q_2}$ in the weak and strong quartic coupling $\frac{\kappa}{(4\pi)^2}$ regimes.

\section{Symmetries and spectrum}%
\label{symmental}
The first step towards understanding the spectrum of fluctuations around our semiclassical solution consists in describing the corresponding symmetry-breaking pattern.

The (Euclidean) Abelian Higgs model with \(N\) complex scalar fields at the \ac{ir} fixed point has symmetry \(SO(d, 1) \times SU(N)/\setZ_{N}\).
The insertion of the cusp, with its vertex sitting in the completely symmetric representation of \(SU(N)\) with \(Q_1 - Q_2\) boxes, breaks the conformal group to the \(SO(d-2)\) rotations orthogonal to the plane of the cups times the dilatations \(\mathcal{D}\), and the global \(SU(N)/\setZ_N\) to \(SU(N-1) / \setZ_{N-1}\).
All in all,
\begin{equation}
  SO( d , 1) \times SU(N)/\setZ_N \to SO( d - 2) \times \mathcal{D}  \times SU(N-1) / \setZ_{N-1} \, .
\end{equation}
In the cylinder frame, the symmetries are realized geometrically: \(SO(d -2)\) corresponds to the isometries of the \(d + 1\) sphere with two non-antipodal punctures, and \(\mathcal{D} \) generates the time translations.

In order to count the Goldstone fields resulting from the breaking of the global symmetry, we observe that in the case of only the global symmetry, the breaking pattern would be the same as for the \(O(2N)\) model: \(SO(d,1) \times O(2N) \to SO(d) \times \mathcal{D}' \times U(N-1)\), where \(\mathcal{D}' \) is the helical symmetry obtained as the sum of time translation and a global \(U(1)\).
In this case, the low-energy theory contains \(1\) type-I Goldstone with linear dispersion relation, \(N - 1\) type-II Goldstones with quadratic dispersion, and \(N-1\) massive modes with mass proportional to \(\mu_0\).
In the presence of the gauge symmetry, the Higgs mechanism comes into play: the type-I Goldstone combines with the photon to give a mass to the vector field. 
The remaining massless fields are the type-II Goldstones with parabolic dispersion relations~\cite{Watanabe:2013uya, Alvarez-Gaume:2016vff, Gaume:2020bmp}.
The fundamental difference with respect to the large-charge analysis of systems with only a global symmetry is that the universal conformal type-I Goldstone is not present here due to the Higgs mechanism~\cite{Antipin:2022hfe}.
For generic values of \(q_1\), \(q_2\) and \(\alpha_{*}\), we thus have a massive spectrum and do not expect to see fluctuations with unit energy (in the scale of \(R\)) to be identified with \ac{cft} descendants of the lowest operator of fixed charge.

To obtain the quadratic Lagrangian, we expand our fields around the classical trajectory.
Even though we have introduced the gauge-invariant field $B$, it is technically simpler to go back to the fields \(A\) and \(\chi\) and introduce the corresponding fluctuations:
\begin{align} \label{flucchio}
  \chi &= - i \mu_0 \tau + \Pi/r \,, & A_\tau &= \frac{i(\mu_0-\mu)}{\sqrt{G}} + \mathcal{A}_\tau  \,, & A_i &=   \mathcal{A}_i + \nabla_i h \,, & \rho_c &= r + \hat{r} \,,
\end{align}
where $\nabla^i \mathcal{A}_i=0$ . The price to pay is the introduction of a gauge fixing term, 
\begin{equation} \label{GF}
  \mathcal{L}_{\xi} = - \frac{1}{2 \xi} \left( \nabla^\mu A_\mu + \sqrt{G} r \Pi \right)^2,
\end{equation}
knowing that the final result has to be gauge invariant. %

In writing Eq.~\eqref{flucchio} we have separated the time component of the gauge field and performed the Helmholtz decomposition for the spatial directions.
Adopting the convenient choice $\xi=1$, the quadratic Lagrangian decomposes into five independent pieces: 
\begin{equation}
  \mathcal{L}^{(2)} =\mathcal{L}_S^{(2)}+\mathcal{L}_T^{(2)}+\mathcal{L}_h^{(2)}+\mathcal{L}_{\text{gh}}^{(2)}+\mathcal{L}_{II}^{(2)}\,.
\end{equation}
$\mathcal{L}_S^{(2)}$ describes the mixing between the $\hat{r}$, $\Pi$, $A_\tau$ scalar modes underlying Higgs mechanism,
\begin{multline}
  \label{Lquad1}
  \mathcal{L}_S^{(2)} = \frac{1}{2}A_{\tau}\left(-\nabla^2+G \ r^2\right)A_{\tau} 
  +\frac{1}{2}(\del_{\mu}\hat{r})^2+\frac{1}{2}(\del_{\mu}\Pi)^2+\frac{1}{2} \left(m_d^2-\mu^2+3 \kappa r^2  \right)\hat{r}^2\\
  -\frac{G r^2}{2}\Pi^2-2i\mu \,\hat{r}\del_{\tau}\Pi+2i \mu r  A_\tau \hat{r}   
   \,,   
\end{multline}
where $m_d = (d-2)/(2 R)$ is the mass stemming from the conformal coupling to the curvature.
$\mathcal{L}_T^{(2)}$ describes the transversal photon polarization,
\begin{equation}
  \label{Lquad2}
  \mathcal{L}_T^{(2)} =  \frac{1}{2} \mathcal{A}_i g^{ii} \left(-\nabla^2 +\frac{d-2}{R^2} +G r^2 \right) \mathcal{A}_i \,,
\end{equation}
where the second term in brackets arises from the commutator of covariant derivatives on $S^{d-1}$, $[\nabla^{i} , \nabla^{j}]=\frac{d-2}{R^2}\delta^{ij}$. 
For the scalar fluctuation $h$, we find the quadratic Lagrangian
\begin{equation}
  \label{Lquadh}
  \mathcal{L}_h^{(2)} =  \frac{1}{2}h\left(-\nabla^2+G \ r^2\right)h \,.
\end{equation}
The inclusion of the gauge fixing term~\eqref{GF} leads to the presence of complex ghost fields $c$ and $\bar c$ with Lagrangian
\begin{equation}
  \mathcal{L}_{\textrm gh}^{(2)} = \bar c \left(\nabla^2 + G r^2 \right)c\,.
\end{equation}
Finally, $\mathcal{L}_{II}^{(2)}$ describes the fluctuations of the scalar fields that do not form a condensate:
\begin{equation}
  \label{Lquad4}
  \mathcal{L}_{II}^{(2)} = \sum_{a=2}^N \left [\del \phi_a^\dagger \del \phi_a+\left( m_d^2 - \mu^2 +\kappa  r^2  \right)\phi_a^\dagger \phi_a  -2 \mu \left(\phi_a^\dagger \del_\tau \phi_a -\phi_a \del_\tau \phi_a^\dagger  \right) \right]\,.
\end{equation}
Due to the nontrivial angular dependence of the classical solution, the calculation of the dispersion relations and the functional determinant of the fluctuation is not straightforward. To make progress, we focus on the small-$G$ limit and determine the $\Gamma^{(0,0)}_{q_1q_2}$ contribution to the cusp anomalous dimension as well as the spectrum to order $G$.
We start by writing the inverse propagator for $\hat{r}$, $\Pi$, and $A_\tau$, in $ \mathcal{L}_S^{(2)}$ as
\begin{align}
{\cal D}_S^{-1} &=\left(
  \begin{array}{ccc}
   - \omega^2+\Delta_{S^{d-1}}+m_d^2-\mu^2+3 \kappa r^2 & -2i\mu\omega   &-2i \sqrt{G}\mu r \\
  2i\mu \omega & -\omega^2+\Delta_{S^{d-1}}+G r^2 &0 \\
   -2i  \sqrt{G} \mu r  & 0 & -\omega^2+\Delta_{S^{d-1}}+G r^2 \\
    \end{array}
\right) \nonumber \\ &= V_0 + \sqrt{G} V_1 + G V_2 +\order{G^{3/2}} \,,  
\end{align}
where $\Delta_{S^{d-1}}$ is the Laplacian on $S_{d-1}$ and
\begin{align}
  V_0 &=
        \begin{pmatrix}
          J_\ell^2+2 \mu_0^2-2 m_D^2-\omega ^2 & -2 i \mu_0 \omega  & 0 \\
          2 i \mu_0 \omega  & J_\ell^2-\omega ^2 & 0 \\
          0 & 0 & J_\ell^2-\omega ^2          
        \end{pmatrix} \,,  \\
  V_1 &=
        \begin{pmatrix}
          0 & 0 & -\frac{i \sqrt{2} \sqrt{ \mu_0(q_1-q_2)}}{\pi R^{3/2} } \\
          0 & 0 & 0 \\
          -\frac{i \sqrt{2} \sqrt{ \mu_0(q_1-q_2)}}{\pi R^{3/2} } & 0 & 0           
        \end{pmatrix} \,,  \\
  V_2 &=
        \begin{pmatrix}
          6\pi   r_1 \left(\mu_0^2-m_d^2\right) \sqrt{\frac{2 R\mu_0}{q_1-q_2}}-2 \mu_0 \mu_1 & -2 i \mu_1 \omega  & 0 \\
          2 i \mu_1 \omega  & \frac{(q_1-q_2)}{2 \pi ^2 R^3\mu_0} & 0 \\
          0 & 0 & \frac{q_1-q_2}{2 \pi ^2 R^3 \mu_0}           
        \end{pmatrix} \, .
\end{align}
In $V_0$ we substituted $\Delta_{S^{d-1}}$ with its eigenvalues $J_\ell^2 = \frac{\ell (\ell+d-2)}{R^2}$ which have multiplicity 
\begin{equation}
  n_\ell = \frac{(2 \ell+d-2) \Gamma (\ell+d-2)}{\Gamma (d-2) \Gamma (\ell+1)}\,.
\end{equation}
The corresponding eigenfunctions are the $Y_{\ell j m}$ spherical harmonics.
We can then determine the three eigenvalues $\lambda_{S,i}$, $i=-,+,3$ of $\mathcal{D}_S^{-1}$ to order $\order{G}$ using perturbation theory.
Denoting the eigenvalues and eigenvectors of $V_0$ as $\lambda^{(0)}_{S,i}$ and $v_{S,i}$, respectively, we have
\begin{equation}
  \lambda_{Si} = \lambda^{(0)}_{S,i} +G \left( \braket{v_{S,i} | V_2 | v_{S,i}} + \sum_{j\neq i} \frac{ \abs{\braket{v_{S,i} |V_1| v_{S,i}}}^2 }{\lambda^{(0)}_{S,i} - \lambda^{(0)}_{S,j}} \right) + \order{G^2}\,.
\end{equation}
The treatment of the other fluctuation modes is analogous. We provide the full result for the eigenvalues in Appendix~\ref{spettro}.
The mass spectrum is obtained by determining the value of $\omega$ that makes the eigenvalues vanish. %
In obvious notation, we obtain 
\begin{align}
    m^2_{S,-} &= G \frac{q_1-q_2}{2 \pi ^2 R^3 \mu_0} + \order{G^2}\,, 
     & m^2_{S,+} = G \frac{q_1-q_2}{2 \pi ^2 R^3 \mu_0} + \order{G^2}\,, \\
        m^2_{S,3} &= 6 \mu_0^2-2 m_d^2 + \order{G^2}\,, 
        & m^2_T = (d-2)+G \frac{q_1-q_2}{2 \pi ^2 R^3 \mu_0} + \order{G^2} \,, \\ 
        m^2_{h} &= 6 \mu_0^2-2 m_d^2 + \order{G^2} \,,
      &  m^2_{\textrm{gh}} =  G \frac{q_1-q_2}{2 \pi ^2 R^3\mu_0} +\order{G^2}\,, \\
      m^2_{II,1} &= \order{G^2}\,, &  m^2_{II,2} = \mu_0^2 + \order{G^2}\,.
\end{align}
As expected, due to the Higgs mechanism, both the relativistic Goldstone boson (described by $\lambda_{S,-}^{(0)}$) present for $G=0$ and the gauge field components acquire a mass for generic values of the charges. This is important because in the standard large-charge \ac{eft} for systems with global symmetry, the spectrum of the Goldstone contains the \ac{cft} descendants, which here are missing.
The exception is the case $q_1=q_2$, for which the scalar condenses and a gapless mode is expected for $G<2\pi$. 

The sum over the order $\order{G}$ eigenvalues presented in Appendix~\ref{spettro} yields the $\Gamma^{(0,0)}$ and  $\Gamma^{(0,1)}$ contributions to the cusp anomalous dimension.
  Here we limit ourselves to discussing $\Gamma^{(0,0)}_{q_1q_2}$ which, by construction, matches the \ac{nlo} semiclassical contribution to the lowest operator with charge $Q_1-Q_2$ at the Wilson-Fisher fixed point of the $O(2N)$ $\lambda \phi^4$ theory.
  We refer the interested reader to references~\cite{Badel:2019oxl,Antipin:2020abu} for the details of the calculation.
The final result is
\begin{equation}
 \Gamma^{(0,0)}_{q_1q_2} = \frac{8 \sqrt{6  (R\mu_0)^2-2}-3 (R\mu_0)^4 (N+4)-6 (R\mu_0)^2 N+16 R \mu_0 (N-1)-7 N+12}{16} + \frac{1}{2} \sum_{\ell=1}^\infty  \sigma (\ell) \,,
\end{equation}
with 
\begin{multline}
    \sigma(\ell) =
  (1-N)\Bigg( 2 \ell^3+6 \ell^2+\left((R\mu_0) ^2+1\right)+ \left((R\mu_0)^2+5\right) \ell-2 (\ell+1)^2 \sqrt{(R\mu_0) ^2+\ell (\ell+2)}\\
  -\frac{\left((R\mu_0) ^2-1\right)^2}{4\ell}  \Bigg) 
  - 2 (\ell + 1)  (R\mu_0)^2 
                 +\frac{5}{4 \ell} \left((R\mu_0)^2-1\right)^2
                 + R (\ell+1)^2 \left(\omega_{+} + \omega_{-}\right) 
                 \,\,,
\end{multline}
where
 \begin{equation}
     \omega_{\pm} = \frac{1}{R}\sqrt{3 (R\mu_0 )^2+\ell (\ell+2) -1
       \pm\sqrt{9 (R\mu_0)^4+(4 \ell (\ell+2)-6)(R \mu_0)^2+1}},
\end{equation}
are the dispersion relations stemming from the eigenvalues $\lambda_{S,\pm}^{(0)}$ of $V_0$ in $d=4$. In the small $\kappa$-limit, the expression above becomes
\begin{multline}
  \label{eq:Gamma00-expansion-small}
  \Gamma_{q_1q_2}^{(0,0)} = -\frac{N+5}{3} \frac{6 \kappa}{(4\pi)^2} \pqty*{q_1 - q_2} +\frac{3-N}{9} \pqty*{\frac{6 \kappa}{(4\pi)^2} \pqty*{q_1 - q_2}}^2 \\
  +\frac{2}{27} (2 (N+7) \zeta(3) + N - 18) \pqty*{\frac{6 \kappa}{(4\pi)^2} \pqty*{q_1 - q_2}}^3 \\
  -\frac{2}{81} \pqty*{(12 N+65) \zeta(3) + 5 (N+15) \zeta(5) + 4 N - 146 } \pqty*{\frac{6 \kappa}{(4\pi)^2} \pqty*{q_1 - q_2}}^4 + \dots  \,.
\end{multline}

\section{Observables}
\label{sec:cusp-dimension}

After the preceding discussions, we are now ready to put everything together and present our final result.
The cusp anomalous dimension is
\begin{multline}
   \Gamma_{q_1 q_2}(\alpha_{*}) = Q \Bigg[ \Gamma_{q_1 q_2}^{(-1, 0)} - \frac{3 \left(q_1 - q_2 \right)^2 + 4 q_1 q_2 \left( 1 + \left( \pi - \alpha_{*} \right) \cot(\alpha_{*}) \right)}{16 \pi^2} G \\
  + \Gamma_{q_1 q_2}^{(-1,2)}( \alpha_{*}) G^2  + \order{G^3} \Bigg] + \Gamma_{q_1 q_2}^{(0,0)} + \order{Q^0 G,1/Q}\,,
\end{multline}
where \(\Gamma^{(-1,0)}\), \(\Gamma^{(-1,2)}\), and \(\Gamma^{(0,0)}\) depend on the double-scaling coupling \(\kappa\). 
Their explicit expressions were given in the previous sections as expansions in \(\kappa \) and \(1/\kappa\) (see Eq.~\eqref{eq:Gamma-10-expansions},~\eqref{eq:Gamma-12-expansion-small}--\eqref{eq:Gamma-12-expansion-big}, and~\eqref{eq:Gamma00-expansion-small}) and in Appendix~\ref{sec:explicit} for arbitrary values of $\kappa$. 

A number of interesting physical quantities can be extracted from the cusp anomalous dimension.
Some of them were discussed in~\cite{Cuomo:2024psk}, where the authors have made some general predictions for generic \acp{cft}, which we are in a position to verify and make precise for the Abelian Higgs model.

\subsection{Shallow cusp and defect-changing operator}

In the limit \(\alpha_{*} \sim \pi\), the cusp is close to the case of the straight infinite Wilson line. The cusp dimension can therefore be reconstructed in terms of the defect-changing operator \(\cO_{Q_1 Q_2}\) and the displacement operator \(D\), a  protected defect operator stemming from the breaking of translational invariance by the line~\cite{Billo:2013jda}.
In this case, we have the expansion
\begin{equation}\label{eq:piexp}
  \Gamma_{q_1 q_2}(\alpha_{*}) = \Delta_{q_1 q_2} + \frac{1}{2} \beta_{q_1 q_2} \pqty*{\alpha_{*} - \pi}^2 + \order{(\alpha_{*} - \pi)^3}\,,
\end{equation}
where \(\Delta_{q_1 q_2}\) is the dimension of the scalar defect operator \(\cO_{q_1 q_2}\) inserted at the origin, and \(\beta_{q_1 q_2}\) depends on the four-point function with two insertions of \(\cO_{q_1 q_2}\) and two insertions of \(D\).
Applied to the Abelian Higgs model, we find for small \(\kappa\)
\begin{align}
  \Delta_{q_1 q_2} ={}& \begin{multlined}[t]
    Q \Bigg[\pqty*{\pqty*{q_1 - q_2}+\frac{\pqty*{q_1 - q_2}^2 \kappa }{8 \pi ^2}-\frac{\pqty*{q_1 - q_2}^3 \kappa ^2}{32 \pi ^4}+ \dots} 
    - \frac{3 \pqty*{q_1 - q_2}^2}{16 \pi ^2} G \\
    +  \Bigg( \tfrac{\left(3 - 8 \pi ^2\right) \pqty*{q_1 - q_2}^3 - 24 \pi ^2 q_1 q_2 \pqty*{q_1 - q_2}}{384 \pi ^4} 
    + \tfrac{\kappa  \left(15 \pi ^2 \left(\pi ^2-6\right) q_1 q_2 \pqty*{q_1 - q_2}^2+2 \left(45-15 \pi ^2+2 \pi ^4\right) \pqty*{q_1 - q_2}^4\right)}{5760 \pi ^6} \\
    +\tfrac{\kappa ^2 \left(120 \pi ^2 \left(\pi ^2-18\right) q_1 q_2 \pqty*{q_1 - q_2}^3+\left(765-720 \pi ^2+32 \pi ^4\right) \pqty*{q_1 - q_2}^5\right)}{184320 \pi ^8}  +  \dots \Bigg) G^2 + \dots \Bigg]\\ -\left[\tfrac{(N+5) (q_1-q_2)\kappa}{8 \pi ^2}+\tfrac{ (N-3) (q_1-q_2)^2\kappa ^2}{64 \pi ^4}+ \dots\right] +\order{Q^0 G} \,,
  \end{multlined} \\
  \beta_{q_1 q_2} ={}& 
                   -\frac{Q G}{6 \pi ^2} + \left(\tfrac{\left(3+\pi ^2\right) \pqty*{q_1 - q_2}}{72 \pi ^4} - \tfrac{\pqty*{q_1 - q_2}^2 \kappa }{32 \pi ^6} + \tfrac{\left(180-105 \pi ^2+14 \pi ^4\right) \pqty*{q_1 - q_2}^3 \kappa ^2}{34560 \pi ^8}+ \dots \right) q_1 q_2 Q G^{2} \nonumber \\ &  +\order{Q G^3,Q^0 G}  \,,                  
\end{align}
and for large \(\kappa\),
\begin{align}
  \label{eq:delta-Q1-Q2}
  \Delta_{q_1 q_2} ={}& \begin{multlined}[t]
    Q \Bigg[\pqty*{\frac{3}{4}\left(\frac{\kappa  (q_1-q_2)^4}{2 \pi ^2}\right)^{1/3}+\dots} 
    - \frac{3 \pqty*{q_1 - q_2}^2}{16 \pi ^2} G \\
      -  \pqty*{\frac{1}{64}\left(\left(3+4 \pi ^2\right) (q_1-q_2)^2+12 \pi ^2 q_1 q_2\right) \left(\frac{(q_1-q_2)^2}{4 \pi ^{10} \kappa }\right)^{1/3} +\dots}G^2 + \dots \Bigg]  \\
  +\bigg[\tfrac{1}{192}\Big(4 (N+4) \log ( (q_1-q_2)\kappa ) -4 (3 \sqrt{3}+13+8 \log (4 \pi )-15 \coth ^{-1}\sqrt{3}) \\-N (15+8 \log (4 \pi ))  +12 \gamma_E  (N+4) \Big)\left(\frac{\kappa ^4 (q_1-q_2)^4}{2 \pi ^8}\right)^{1/3}+ \dots\bigg] +\order{Q^0 G}\,,
  \end{multlined} \\
  \beta_{q_1 q_2} ={}& 
                   -\frac{Q G}{6 \pi ^2} + \left(\frac{3}{16}\left(\frac{(q_1-q_2)^2}{4 \pi ^{10} \kappa }\right)^{1/3}+\dots\right) q_1 q_2 Q G^{2} +\order{Q G^3,Q^0 G} \,.                  
\end{align}
The explicit expressions for arbitrary values of $\kappa$ can be found in App.~\ref{sec:explicit}.

The dimension \(\Delta_{q_1 q_2}\) vanishes for \(q_1 = q_2=q\), as it should since in this case \(\cO_{q q}\) is the identity.
Furthermore, the whole \(\order{G^2}\) contribution vanishes and
\begin{equation}
  \beta_{q q} = -\frac{Q G}{6 \pi ^2} +\order{Q G^3,Q^0 G}    = -\frac{(Q e)^2}{6 \pi^2} + \order{Q e^6,Q^0 e^2}\,,
\end{equation}
which is negative as demanded by reflection positivity.

\subsection{Sharp cusp and superconducting order parameter}
\label{sec:sharpschwinger}

In the opposite limit, \(\alpha_* \to 0\), $\Gamma_{q_1 q_1}$ describes the fusion of the two antiparallel half-lines of charge \(q_1\) and \(q_2\).
The form of the anomalous dimension is set by the defect fusion algebra to be
\begin{equation}\label{eq:defectdim}
  \Gamma_{q_1 q_2}
  = \frac{C_{q_1 (-q_2) (\overline{q_1 - q_2})}}{\alpha_*}
  + \Delta_{(q_1 - q_2) 0}
  + \alpha_{q_1 q_2}\, \alpha_*^{\Delta_{\text{irr}}-1}
  + \ldots \, .
\end{equation}%
Here, \(C_{q_1 (-q_2) (\overline{q_1 - q_2})}\) is the Casimir interaction between the two conformal lines at separation of order \( \order{\alpha_*}\) and measures the energy cost required to fuse them into a single line of charge \(q_1 - q_2\) (binding energy).
Since all the components of \(\Gamma_{q_1 q_2}\) aside from \(\Gamma^{(-1,1)}_{q_1q_2}\) are regular at \(\alpha_{*} = 0\), we find
\begin{equation}
  C_{q_1 (-q_2) (\overline{q_1 - q_2})} = - \frac{G Q}{4 \pi} q_1 q_2 = - \frac{e^2 }{4 \pi} Q_1 Q_2 \, ,
\end{equation}
which, as expected, is the Coulomb potential between the two charges.
We are also able to extract the dimension of the lowest irrelevant operator on the $\bar Q$ line, which is $\Delta_{\text{irr}} = 2 +\order{G^3}$ as well as the associated coefficient $\alpha_{q_1q_2}$, which reads
\begin{equation}
  \alpha_{q_1q_2} = \frac{ Q G q_1 q_2}{12 \pi }\left( 1 + \tfrac{\frac{3^{5/3}}{4 \pi^{4/3}} \left(\sqrt{81 \kappa ^2 (q_1 - q_2)^2-48 \pi ^4}+9 \kappa  (q_1 - q_2)\right)^{1/3} (q_1-q_2)}{2 ( 6 \pi^4)^{1/3} +\left(\sqrt{81 \kappa ^2 (q_1 - q_2)^2-48 \pi ^4}+9 \kappa  (q_1 - q_2)\right)^{2/3}} G + \order{G^2} \right) + \order{Q^0 G} \,.
\end{equation}

From the finite contribution of $\Gamma_{q_1 q_2}$ in the limit $\alpha_* \to 0$, we obtain the dimension of the creation operator $\phi_{(i_1}\dots \phi_{i_{\bar Q})}$ of the fused defect, which is just a half line with charge $q_1-q_2$. 
Alternatively, we can think of it as a semi-infinite line joining two scalar insertions of opposite charges.
This is the usual \ac{ms} dressed two-point function $\braket{\cO_{\bar{Q}}^{*}(x_\infty) \,W_{\gamma}(\bar{Q}) \, \cO_{\bar{Q}}(0)}$.
Its scaling at criticality, which is used as an order parameter for superconductivity, is manifestly the dimension of the defect-creation operator \(\Delta_{\bar Q 0}\): we can think of the \ac{ms}-dressed two-point function as the (regularized) limit of the cusp either for  \(\alpha_{*} \to 0\), or for \(q_2 \to 0\).

Our \(\order{G^2}\) result falsifies a standing conjecture about the scaling of the \ac{ms} dressing~\cite{Kleinert_2003,Kleinert:2005sa}.
While expectation values of non-gauge invariant operators vanish identically, one can still compute perturbatively a ``gauge-dependent'' conformal dimension for the scalar field.
It has been observed that if one takes this quantity and imposes the traceless gauge, the \(\order{G}\) result coincides with the scaling of the gauge-invariant \ac{ms} two-point function.
Equation~\eqref{eq:delta-Q1-Q2} shows explicitly that this is accidental and does not generalize to higher orders in \(G\).
The story is different for the Dirac dressing that we review in Appendix~\ref{sec:Dirac}, since in this case the (physically meaningful) scaling dimension coincides with the gauge-dependent result in Landau gauge.

\subsection{Trivial defect-changing operator}

In the special case \(Q_1 = Q_2\), the \(\order{G^2}\) term vanishes and the cusp anomalous dimension reduces to
\begin{equation}
  \label{eq:trivial-cusp}
  \Gamma_{qq}(\alpha_{*}) = - Q \frac{1 + \pqty{\pi - \alpha_{*}} \cot(\alpha_{*})  }{4 \pi^2} G + \order{G^3}\,,  
\end{equation}
and, consistently with reflection positivity,
\begin{align}
  \Gamma_{qq}(\alpha_{*}) & < 0\,, &  \Gamma'_{qq}(\alpha_{*}) & > 0\,, &  \Gamma''_{qq}(\alpha_{*}) & < 0 \, .
\end{align}
From the second derivative in \(\alpha_{*}\) we can derive the (physically meaningful) normalization of the displacement operator:
\begin{equation}
  \braket{ D_i(\tau) D_j (0) } = - \frac{6  \Gamma''_{qq}(\pi)}{\tau^4} = \frac{Q G}{\pi^2} \frac{1}{\tau^4} = \frac{e^2 Q^2}{\pi^2} \frac{1}{\tau^4} \,.
\end{equation}
In fact, this result can be made stronger: if \(q_1 = q_{2}\), there is no scalar insertion at the cusp and so, at least when $G$ is sufficiently small, the minimum-energy configuration is characterized by a vanishing classical profile for $\rho_c$. This allows us to solve the Maxwell equations exactly, obtaining
\begin{equation} \label{equalcharges}
  \mu(\alpha, \gamma) = \frac{G}{4 \pi ^2 R}\left((\pi -\alpha ) \cot (\alpha )-(\pi -\gamma ) \cot (\gamma )\right) \, .
\end{equation}
This means that in our regime, the value that we found in Eq.~\eqref{eq:trivial-cusp} does not receive perturbative corrections in \(G\) at all and so $\Gamma_{qq}^{(-1)}=G \Gamma_{qq}^{(-1,1)}$.
On the one hand, this matches the known one-loop cusp anomalous dimension in Maxwell theory~\cite{Korchemsky:1987wg}.
On the other, we find an all-order statement from the point of view of conventional perturbation theory: in the perturbative expansion of the cusp anomalous dimension, all the terms scaling as $Q_1 G^i \kappa^k = Q_1^{1+i+k} e^{2i}\lambda^k$ vanish except the one with $i=1 \,, \ k=0$.

One may be tempted to conclude that the scalar fields do not contribute to the cusp anomalous dimension in the semiclassical limit \emph{i.e.}, in the case of large charge Wilson lines. However, as discussed earlier, for $G> 2\pi$ one generally expects the path integral to be dominated by a new trajectory characterized by a nontrivial scalar profile as shown in~\cite{Aharony:2023amq} for the straight Wilson line ($\alpha_*=\pi$) case.

\section{Perturbative \(\epsilon\)-expansion}
\label{sec:pert-epsil-expans}

As mentioned earlier, while our results are valid for any fixed value of \(\kappa\), the limit of small $\kappa$ overlaps with the regime of validity of standard (Feynman diagram) perturbation theory in \( 4 - \epsilon\) dimensions.

This provides a controlled semiclassical expansion even when the underlying couplings are not parametrically small. In this sense, large charge offers an alternative organizing principle, complementary to weak coupling, that allows one to access nontrivial regimes of Wilson line observables and defect dynamics.

To compare the two approaches, we can reinterpret our semiclassical construction as a rearrangement of the terms in the conventional \(\epsilon\)-expansion.
In Eq.~(\ref{eq:cusp-rescaled-action}) we have rewritten the action in the form \(S = Q \bar S\), where \(\bar S\) depends only on the 't Hooft couplings, obtained by keeping the product \(Q \epsilon\) fixed.
Now \(Q\) is a loop-counting parameter and, order-by-order in \(Q\), the observables are functions of the product \(\epsilon Q\). 

Concretely, when expanded in the limit of small \(\epsilon Q\), the leading-order (semiclassical) result $ \Gamma_{Q_1 Q_2}^{(-1)}$ yields all terms scaling as \(Q (Q \epsilon)^n\), the one-loop result $ \Gamma_{Q_1 Q_2}^{(0)}$ the terms scaling as \(Q^0 (\epsilon Q)^n\), and in general the \(k\)-loop contribution $ \Gamma_{Q_1 Q_2}^{(k-1)}$ in the double-scaled action around the semiclassical saddle provides a resummation of all terms scaling as \(Q^{1-k}(\epsilon Q)^n\).
This is to be compared with the standard \(\epsilon\) expansion, where at each order in \(\epsilon\) there will be terms that scale with non-negative powers of \(Q\) (since the \(Q \to 0\)  limit is regular), showing that the two expansions --- small \(\epsilon\) and small \(\epsilon Q\) --- are complementary.
Rewriting the 't Hooft couplings in terms of the original \(e^2\) and \(\lambda\), and expanding at second order in \(\epsilon\) (remembering that \(\lambda = \order{\epsilon}\) and \(e = \order{\sqrt{\epsilon}}\)), we find for the cusp anomalous dimension the following expansion:
\begin{equation}
  \label{smalllambda}
  \begin{aligned}
    \Gamma_{Q_1 Q_2}(\alpha_{*}) ={}& Q_1 - Q_2  + \frac{ \lambda}{3}  (Q_1-Q_2) (Q_1-Q_2-5-N)  \\
    &-\frac{e^2}{16 \pi ^2}\left( 3 (Q_1-Q_2)^2+4 Q_1 Q_2 \left (1+ (\alpha_*-\pi ) \cot \alpha_* \right) + \order{Q_1-Q_2}\right)   \\ %
    &
      \begin{multlined}[b][.85\textwidth]
        - \frac{\lambda ^2}{9}  \left(2 (Q_1-Q_2)^3 +(N-3) (Q_1-Q_2)^2 + \order{Q_1-Q_2}\right) \\
        + \order{\lambda e^2(Q_1-Q_2)^2, \lambda e^2 Q_1 Q_2}
      \end{multlined}
    \\
                          &
                            \begin{multlined}[b][.85\textwidth]
                              + \frac{e^4 (Q_1-Q_2)}{384 \pi ^4}  \Big( \left(3-8 \pi ^2\right) (Q_1-Q_2)^2 \\
                              + 8 Q_1 Q_2 \left(\alpha_* (\alpha_*-2 \pi ) (\alpha_*-\pi ) \cot \alpha_*-2 \pi ^2\right) + \order{Q_1-Q_2}\Big) + \dots\,,
                            \end{multlined}
  \end{aligned}
\end{equation}
where the dots denote $\order{\epsilon^3}$ corrections. In the $Q_2 = 0$ case, we recover the one-loop result found in~\cite{Kleinert_2003, Antipin:2022hfe} for the scaling dimension $\Delta_{Q_10}$ associated with the \ac{ms} dressed two-point function and make a prediction for a two-loop calculation.

\section{Conclusions}
\label{sec:conclusions}

When applying the semi-classical methods of the large-charge expansion to gauge theories, the need to work with gauge-invariant quantities leads directly to the notion of nonlocal operators, and particularly those constructed from Wilson lines.
A natural set-up is to consider two Wilson lines meeting at a cusp, with the two arms carrying different charges $Q_1$, $Q_2$.
This case encompasses both the semi-infinite line, and the straight Wilson line without cusp but with a local operator insertion, as limiting cases.

From the viewpoint of defect quantum field theory, a Wilson line defines a one-dimensional defect, and the large-charge limit probes sectors where it carries large quantum numbers, in close analogy with the standard large-charge expansions in \ac{cft}.
This framework provides an alternative organizing principle for gauge theories, complementary to weak coupling, and enables a systematic study of non-linear and backreaction effects in Wilson line observables.

In this paper, we have focused on the Abelian Higgs model in $d=4-\epsilon$ dimensions.
We have successfully studied two Wilson lines carrying charges $Q_1$, $Q_2$ with a defect-changing operator at the cusp.
To be able to work semiclassically, we have focused on the double-scaling limit in which $Q_1$, $Q_2$ are large and $Q\epsilon$ is constant.
Despite taking this limit, we still needed to work perturbatively in the gauge-coupling $G$ (but for finite values of the self-coupling \(\kappa\)).
The semiclassical computation of the cusp anomalous dimension to \ac{nnlo} in $G$ constitutes our main result.
Schematically, we found
\begin{multline}
   \Gamma_{q_1 q_2}(\alpha_{*}) = Q \Bigg[ \Gamma_{q_1 q_2}^{(-1, 0)} - \frac{3 \left(q_1 - q_2 \right)^2 + 4 q_1 q_2 \left( 1 + \left( \pi - \alpha_{*} \right) \cot(\alpha_{*}) \right)}{16 \pi^2} G \\
  + \Gamma_{q_1 q_2}^{(-1,2)}( \alpha_{*}) G^2  + \order{G^3} \Bigg] + \Gamma_{q_1 q_2}^{(0,0)} + \order{Q^0 G,1/Q}\,,
\end{multline}
where all the functions are known explicitly and are given above.
Our results reproduce those of the standard perturbative calculation in the literature while greatly extending the accessible regimes, since they are valid for \(Q_1 \neq Q_2\) and for large values of the charges.

There are a number of directions one could follow from here.
\begin{itemize}
\item The small-coupling expansion that we have used breaks for \(G \approx 2 \pi\).
  This points to a new phase transition (analogous to the one discussed in~\cite{Aharony:2022ntz} for a straight Wilson line) that could be discussed with our semiclassical methods.
\item The physics of non-Abelian gauge symmetries is quite different and much richer. Also in that case, a semiclassical analysis might lead to new nontrivial regimes of Wilson line observables and defect dynamics.
\item  It would be interesting to see whether our semiclassical approach would be useful in more general geometric settings involving multiple cusps such as \emph{e.g.} Wilson lines with two or more cusps, or Wilson loops with $n$ cusps forming a polygon, and whether it could describe interactions between cusps.
\item There is a large literature on more general geometries with junctions in which more than two Wilson lines meet, or networks with junctions~\cite{Almelid:2015jia, Kidonakis:1998nf, Dixon:2008gr, Becher:2009cu}.
  It could be worth exploring whether working semiclassically would be useful or could capture interactions between junctions.
\item We have shown that focusing on sectors of the theory where the Wilson lines carry large quantum numbers leads to important simplifications.
  This should naturally extend to higher-dimensional defects seen as semiclassical objects to be studied systematically in a large quantum number expansion.
\end{itemize}

\section*{Acknowledgments}

\begin{small}\sffamily
  The authors would like to thank Oleg Antipin, Lorenzo Bianchi and Marco Meineri for useful discussions and Oleg Antinpin for detailed comments on the manuscript.
  J.B., S.R., and J.W. are supported by the Swiss National Science Foundation under grant number 200021\_219267.
  J.W. would like to thank the Ramsay Centre for Western Civilisation for its support.
\end{small}

\appendix

\section{Dirac dressing}
\label{sec:Dirac}

Besides the cusp studied in the main part of this paper, the Abelian Higgs model contains other non-local operators that at weak coupling take the form of two insertions of the scalar field and an integral of the gauge field.
In fact, every object of the form
\begin{equation}\label{eq:A1}
 \phi(x_1) \exp*[i e \int\dd^4{y} A_{\mu}(y) J^{\mu}(y) ] \phi^{*}(x_2) \,, 
\end{equation}
is gauge invariant if the current \(J_{\mu}\) satisfies the condition
\begin{equation}
  \del_{\mu} J^{\mu}(y) = \delta( y - x_2) - \delta( y - x_1) \, ,
\end{equation}
and its expectation value is a physical observable (by Elitzur's theorem, the expectation value of a gauge-dependent operator is necessarily vanishing~\cite{Elitzur:1975im}).

Standard examples include the Wilson line with endpoints at $x_1$ and $x_2$. For a straight line, it corresponds to the \ac{ms} dressing limit of the cusp discussed in Section~\ref{sec:sharpschwinger}. For generic endpoints, the \ac{ms} currents in Eq.~\eqref{eq:JMS} becomes
\begin{equation}
     J_{\text{MS}}^\mu(x;\gamma)= \pqty{x_1^\mu - x_2^\mu} \int_0^1 \dd{s} \delta^d\lrp{y^\mu - (s x_1^\mu + (1-s) x_2^\mu)} \,.
\end{equation}
Another well-known possibility is the so-called Dirac-dressed propagator~\cite{Dirac:1955uv}, for which \(J^\mu\) is the gradient of the sum of Green's functions \(G_0\) centered at \(y = x_1\) and \(y = x_2\) \footnote{Note that the Dirac dressing can also be expressed as a Wilson line of the longitudinal part $A^L_\mu$ of the gauge field, as considered for example in~\cite{Herbut}.} :
\begin{equation}
 J^{\mu}_{D}(y) = \del^{\mu} \bqty*{ G_0( y - x_2) - G_0( y - x_1)} = \frac{\Gamma(d/2 -1)}{4 \pi^{d/2}} \pdv{}{y^{\mu}} \bqty*{\frac{1}{\abs{y -x_2}^{d/2}} - \frac{1}{\abs{y -x_1}^{d/2}}}   \, .
\end{equation}
This may be seen as the minimal dressing, as any dressing current can be written as 
\begin{equation}
    J^\mu(y,x_1,x_2)=\del^\mu G_0(x_2-y)-\del^\mu G_0(x_1 -y)+h^\mu(y,x_1,x_2), \quad \del_\mu h^\mu=0,
\end{equation}
with the Dirac choice being simply $h^\mu=0$. In this case, one can further ``break'' the exponential dressing factor into two pieces and think of the expectation value as a propagator for a dressed field \(\phi_D\):
\begin{gather}
  \Braket{ \phi(x_1) \exp*[i e \int\dd^d{y} A_{\mu}(y) J_D^{\mu}(y) ] \phi^*(x_2)} = \braket{\phi_D(x_1) \phi_D^*(x_2)} \, , \\
  \phi_D(x) = \phi(x) \exp*[i e \int\dd^d{y} G_0(y - x) \del_\mu A^{\mu}(y)] \, .
\end{gather}
Loosely speaking, one can think of $\phi_D$ as an electron together with the Coulomb field it generates (or, better, a gauge-invariant charged excitation embedded in a screened medium). From the definition of $\phi_D$, it is evident that in the Lorentz gauge \(\del_{\mu} A^{\mu} = 0\), the dressing disappears and $\phi_D$ coincides with $\phi$.  $\phi_D$ realizes a suitable order parameter for the phase transitions described by the Abelian Higgs model as its phase rotation symmetry is broken in the ordered phase, where it acquires a nonzero expectation value~\cite{Kennedy:1986ut,Bonati:2024sok}. Further, the two-point function of $\phi_D$ is proportional to the current measured in the Josephson effect~\cite{Beekman:2019pmi}.

As observed in~\cite{Antipin:2022hfe}, the Dirac-dressed propagator of \(\phi_D^Q\) and the corresponding scaling dimension $\Delta_D$ controlling its scaling at criticality can be studied semiclassically in the same double-scaling limit that we have been using in this work. Borrowing the notation introduced in Eq.~\eqref{semicusp}, the semiclassical $1/Q$ expansion assumes the form
\begin{equation}
  \Delta_D = Q \Delta_D^{(-1)} + \Delta_D^{(0)}  + \frac{1}{Q} \Delta_D^{(1)} + \dots
\end{equation}
In particular, the authors of \cite{Antipin:2022hfe} determined $\Delta_D^{(-1)}$ and $\Delta_D^{(0)}$ as the energy on the cylinder in the presence of a homogeneously charged background and then invoked an argument based on the state–operator correspondence, despite the operator in question being non-local. 
In this appendix, we first compute the two-point function directly in flat space and find that it exhibits the expected power-law behavior, thereby providing explicit support for the arguments presented in the literature and demonstrating the validity of the state-operator correspondence. Moreover, we generalize the calculation to general $N$ and further investigate the strongly-coupled $Q \epsilon \gg 1$ regime.

In order to compute the dimension of the Dirac dressed two-point function of \(\phi_D^Q\), we map the two insertions to the origin and the point at infinity, and consider the analog of the rescaled action in Eq.~(\ref{eq:cusp-rescaled-action-for-B}) which, in the same double-scaling limit, reads
\begin{multline}
  \label{eq:dirac-rescaled-action}
 \bar S  =  \int \dd^d{x} \Big(- \frac{1}{4G}   \tilde F_{\mu\nu}   \tilde F^{\mu\nu} +\frac12 \left(\del \rho\right)^2-\frac12 \rho^2 B^\mu B_\mu+ \frac{\kappa}{4} \rho^4
  -  \log \rho \pqty*{ \delta(x) + \delta(x - x_{\infty}) } - G J_D^\mu B_\mu\Big)  \, .
\end{multline}
The relevant saddle preserves the \(SO(d)\) symmetry and is isotropic.
In the spherical coordinate system of Eq.~(\ref{eq:spherical-coordinates}), the only non-zero component of the Dirac current is in the \(\tau \) direction and is constant:
\begin{equation}%
  J_D^\tau =- \frac{1}{R^{d-1} \Omega _{d-1}}\,.
\end{equation}
The Gauss law yields
\begin{equation}
\rho_c^2 B_\tau =  \frac{1}{R^{d-1} \Omega _{d-1}}\,,
\end{equation}
and the right-hand side of the Maxwell equations vanishes identically.
Physically, \(J_D\) acts as a spatially homogeneous neutralizing background charge density.

The classical solution to the \ac{eom} coincides with the one that we have found in Section~\ref{sec:semiclassical-cusp} for \(G = 0\), and the expectation value of the dressed propagator is
\begin{equation}
  \Braket{ \phi_D^Q(x_1){\phi^*_D}^Q(x_2)}  = \frac{1}{\abs{x_1 - x_2}^{2 Q \Delta_D^{(-1)}}}  \,,
\end{equation}
where \(\Delta_D^{(-1)}\) coincides with the expression of \(\Gamma^{(-1,0)}_{q_1 q_2}\) in Eq.~\eqref{primo} (evaluated for $q_1=1$ and $q_2=0$), in agreement with the results of~\cite{Antipin:2022hfe}. That the state-operator correspondence works can be better understood by noting that the Dirac-dressed two-point function does not break more symmetries than the scalar field insertions already do, \emph{i.e.} it realizes the symmetry-breaking pattern
\begin{equation}
  SO(d,1) \times SU(N)/\setZ_N \to SO(d) \times \mathcal{D} \times SU(N-1)/\setZ_N  \,.
\end{equation}
It can be shown that in four dimensions, demanding conformal invariance of the dressing term in Eq. \eqref{eq:A1} uniquely selects the Dirac current. This can also be seen by  employing the Hodge decomposition on a manifold $\mathcal{M}$ to obtain $A_\mu=\del_\mu a + A_\mu^T \,,$ with $\del^\mu A_\mu^T=0$, to express the dressing factor as
\begin{multline}
    \exp*(i e \int_\mathcal{M} \dd{y}  \del_\mu a(y) \del_\mu G(x-y) )\\
    = \exp*(-i e \int_\mathcal{M} \dd{y} a(y)\Box G(x-y)+i e\oint_{\del \mathcal{M}} d\Sigma_\mu a(y)\del^\mu G(x-y) )\\
    =\exp*(i e \alpha(x)-i e\alpha(\infty) ) \,,
\end{multline}
which appears as two local operator insertions. When the insertion point is mapped to the origin, the configuration preserves the $SO(d)$ isometries of the ($d-1$)-sphere. This also makes manifest how the Dirac dressing measures the local phase of $\phi$ relative to its phase at infinity.

The parallel with the semiclassical calculation in the global case, however, does not extend to the fluctuation spectrum.
Indeed, gauge symmetries
cannot be broken spontaneously.
In perfect analogy with what we have discussed in Section~\ref{symmental}, the would-be conformal Goldstone of the generalized superfluid is Higgsed and the spectrum is massive, apart from the type-II Goldstones.

Since the classical solution is homogeneous, one can determine the dispersion relation of the quantum fluctuations and compute the corresponding functional determinant, yielding the \ac{nlo} contribution to the energy in the semiclassical expansion.
For $N=1$, this has been computed in~\cite{Antipin:2022hfe}.
In order to keep the formulas relatively compact, it is convenient to express the results in terms of the function \(\mu_0 = \mu_0(\kappa(q_1 - q_2))\) in Eq.~\eqref{eq:mu0}.
Then we have
\begin{multline}
  \Delta_D^{(a)}  =\,\, \frac{1}{16} \left(-15 (R\mu_0) ^4-6 (R\mu_0) ^2+8 \sqrt{6 (R \mu_0) ^2-2}+5\right) + \frac{1}{2} \sum_{\ell=1}^\infty  \sigma^{(a)} (\ell)  \\
  -\frac{G}{16\kappa}((R\mu_0)^2-1)\left(\frac{G}{2\kappa}(7 (R\mu_0)^2+5) -9 (R \mu_0)^2 + 5\right) \,,
\end{multline}
where $\sigma^{(a)}(\ell)$ is
\begin{multline}
    \sigma^{(a)}(\ell) =
                 \frac{3G}{4\kappa \ell} \left((R\mu_0)^2-1\right)\left[\left(\frac{G}{2\kappa}-1\right)\left((R\mu_0)^2-1\right)-2 \ell(\ell+1)\right]- 2 (\ell + 1) ( 2 \ell (\ell + 2) + (R\mu_0)^2)  \\ 
                 +\frac{5}{4 \ell} \left((R\mu_0)^2-1\right)^2
                 +R (\ell+1)^2 \left(\omega_{+} + \omega_{-}\right) +2 R\ell (\ell+2) \omega_T
                 \,\,,
\end{multline}
with
{\small \begin{align}
\omega_{\pm} &= \frac{1}{R}\sqrt{\frac{G}{2\kappa}\left((R\mu_0) ^2-1\right)+3 (R\mu_0) ^2+\ell (\ell+2)-1
       \pm\sqrt{\left(\frac{G }{2\kappa}\left((R\mu_0) ^2-1\right)-3  (R\mu_0) ^2+1\right)^2+4 \ell (\ell+2) (R\mu_0) ^2}}\,,
     \\ \omega_T &=\frac1R \sqrt{\frac{G}{\kappa} \left((R\mu_0)^2-1\right) +\ell (\ell+2)+1}\,.
\end{align}}
Here, $\omega_{\pm}$ is the energy of the two physical modes arising from the quadratic Lagrangian $\mathcal{L}_S^{(2)}$ in Eq.~\eqref{Lquad1} evaluated for $\mu=\mu_0$. In the $G \to 0$ limit, $\omega _-$ describes the relativistic Goldstone boson appearing in the large-charge \ac{eft} description. Additionally, $\omega_T$ is the dispersion relation of the physical polarizations of the gauge field with quadratic Lagrangian $\mathcal{L}_T^{(2)}$ in Eq.~\eqref{Lquad2}.
For general $N$, one has to add the functional determinant of the fluctuation fields described by the quadratic Lagrangian $\mathcal{L}_{II}^{2}$ in Eq.~\eqref{Lquad4} with $\mu=\mu_0$, which yields~\cite{Antipin:2020abu}
\begin{align} 
 \Delta_D^{(b)}& =\frac{1-N}{16}\left(3 (R\mu_0) ^4+6 (R\mu_0)^2 -16 R \mu_0+7+16 \sum_{\ell=1}^\infty \sigma^{(b)}(\ell) \right) \,,
\end{align}
  with
\begin{align}
    \sigma^{(b)}(\ell) &= 2 \ell^3+6 \ell^2+\left((R\mu_0 )^2+1\right)+ \left((R\mu_0) ^2+5\right) \ell-2 (\ell+1)^2 \sqrt{(R\mu_0) ^2+\ell (\ell+2)} - \tfrac{\left((R\mu_0) ^2-1\right)^2}{4\ell}\,,
\end{align}
such that the leading semiclassical correction to the Dirac-dressed correlator is given by $ \Delta_D^{(0)}= \Delta_D^{(a)}+ \Delta_D^{(b)}$.

Since the spectrum does not contain massless modes, it is clear that the large-charge behavior of the Dirac-dressed correlator cannot be described by the standard large-charge \ac{eft}. It is therefore of interest to compare the large $Q \epsilon$ behavior of $\Delta_D$ to the general prediction of the latter for the dimension $\Delta_Q$ of the lowest operator with charge $Q$ in models with global $U(1)$ symmetry, which in $d=4-\epsilon$ dimension is~\cite{Hellerman:2015nra}
\begin{equation}
  \label{largen}
  \Delta_Q = \delta_{4/3} Q ^{\frac{4-\epsilon }{3-\epsilon }} +\delta_{2/3} (Q \epsilon )^{\frac{2-\epsilon }{3-\epsilon }} +\delta_0(Q \epsilon )^{-\frac{\epsilon }{3-\epsilon }}  + \order{Q ^{\frac{-2-\epsilon }{3-\epsilon } } }\,.
\end{equation}
By solving Eq.~\eqref{eq:LOparam} in the large $\kappa$ limit, we straightforwardly obtain the large $Q \epsilon$ expansion of $ \Delta_D^{(-1)}$ :
\begin{equation}
  \Delta_D^{(-1)} =\frac{3^{2/3}\lambda^{1/3} Q^{4/3}}{2^{4/3}}+\frac{3^{1/3} Q^{2/3}}{2\ 2^{2/3}\lambda^{1/3}}  -\frac{1}{16 \lambda } +\order{Q^{-2/3}\lambda^{-5/3}}\,.
\end{equation}
Additionally, the large $Q \epsilon$ limit of $\Delta_D^{(0)}$ is obtained by noting that the large $Q \epsilon$ limit corresponds to the limit $R \mu_0 \to \infty$. In turn, the large $R \mu_0$ expansion of $\Delta_D^{(0)}$ can be derived by making use of the Euler--McLaurin formula as explained in~\cite{Cuomo:2021cnb}, obtaining
\begin{equation}
  \tilde \Delta_D = A_4 \pqty*{R\mu_0}^4 + A_2 \pqty*{R\mu_0}^2 + A_0 + A_L\left( \pqty*{R\mu_0}^2 - 1\right)^2 \log( R\mu_0 )\,,
\end{equation}
where 
\begin{align}
    A_L &= \frac{N (N+4)}{4 (N+x+18)} \,,\\
    A_4 &= \tfrac{N (6 \gamma_E  (N-x+18)-7 N+7 x-144)-648 (N-1) \log (2)-36 x-378}{5184}+ \int_0^{\infty } \Sigma_4(k) \, dk \,,\\
    A_2&=\frac{((6 \gamma_E -1) N+45) (-N+x-18)+648 (N-1) \log (2)}{2592} + \int_0^{\infty } \Sigma_2(k) \, dk\,,\\
    A_0 &=\tfrac{N (30 \gamma_E  (N-x+18)+25 N-25 x+1584)-3240 (N-1) \log (2)-270 x+6804}{25920}+ \int_0^{\infty } \Sigma_0(k) \, dk \,,
\end{align}
with $x =\sqrt{(N-180) N-540}$.
The integrand functions appearing above are rather cumbersome,%
\footnote{All explicit expressions can be provided upon request in a Wolfram Mathematica notebook.}
 so we present here only $\Sigma_4$, which reads
\begin{multline}
  \Sigma_4 = \frac{1}{864} k \Bigg(\frac{(N-90-x) N+108}{k^2+1}-36 (N+42-x) +72 k \bigg(2 \sqrt{2} \sqrt{18 k^2+N-x+18} \\
  + \sqrt{36 k^2+N-x+126+\sqrt{2} \sqrt{2592 k^2+N (N-x-180)+90 (x+42)}} -24 k  \\
  +\sqrt{36 k^2+N-x+126-\sqrt{2} \sqrt{2592 k^2+N (N-x-180)+90 (x+42)}}\bigg)\Bigg) \, .     
\end{multline}
Using the above, we find that the large $Q \epsilon$ behavior of $\Delta_D$ matches Eq.~\eqref{largen} with
\begin{align}
     \delta_{4/3} & =\frac{3 A_L^{1/3}}{4\ 2^{2/3}} +\epsilon ^{\frac{1}{3-\epsilon }}\left(\frac{A_L^{-4/3}   \left(3 A_4-A_L \log \left(4 A_L\right)\right)}{12\ 2^{2/3}}\epsilon+\order{\epsilon^2} \right),\\
      \delta_{2/3} & =  \epsilon ^{\frac{1}{-3+\epsilon }}\left(\frac{1}{(2 A_L)^{1/3}}+\frac{ A_L^{-2/3} \left(9 A_2+12 A_4+3 A_L+2 A_L \log \left(4 A_L\right)\right)}{18\ 2^{1/3}}\epsilon  +\order{\epsilon^2} \right),\\
       \delta_0 & = \epsilon ^{\frac{3}{-3+\epsilon }}\left(-\frac{A_L}{3} +  \left(A_0+\frac{2 A_2}{3}+\frac{2 A_4}{3}-\frac{5 A_L}{18}-\frac{1}{9} A_L \log (4 A_L)\right)\epsilon+\order{\epsilon^2} \right),
\end{align}
where higher-order corrections in $\epsilon$ stem from subleading semiclassical orders. We therefore conclude that the large  $Q \epsilon$ behavior of $\Delta_D$ agrees with the generic prediction of the large-charge \ac{eft} despite the absence of gapless modes in the spectrum. 

\section{Some explicit results}
\label{sec:explicit}
In this appendix, we collect our results for $ \Gamma_{q_1 q_2}$ at the leading order in the semiclassical expansion and to various orders in the rescaled gauge coupling $G$. First, we provide the explicit expressions for $\Gamma_{q_1 q_2}^{(-1,0)}$, which is obtained by plugging Eq.~\eqref{eq:mu0} into Eq.~\eqref{primo}:
\begin{multline}
    \Gamma_{q_1 q_2}^{(-1,0)} = \frac{q_1 - q_2}{4}  \Bigg(\left(\tfrac{3(\sqrt{81 \pqty{q_1 - q_2}^2 \kappa ^2-48 \pi ^4}+9 \pqty{q_1 - q_2} \kappa )}{4\pi^2}\right)^{1/3} \\
  +\tfrac{2 \left((6 \pi )^{2/3} \left(\sqrt{81 \pqty{q_1 - q_2}^2 \kappa ^2-48 \pi ^4}+9 \pqty{q_1 - q_2} \kappa )\right)^{2/3}+6 \pi ^2\right)}{2 \left(6 \pi ^4 \left(\sqrt{81 \pqty{q_1 - q_2}^2 \kappa ^2-48 \pi ^4}+9 \pqty{q_1 - q_2} \kappa \right)\right)^{1/3}+\sqrt{81 \pqty{q_1 - q_2}^2 \kappa ^2-48 \pi ^4}+9 \pqty{q_1 - q_2} \kappa }\Bigg) \,.
\end{multline}
Next, the order $\order{G^2}$ contribution to the semiclassical cusp anomalous dimension reads
\begin{equation}
    \Gamma_{q_1 q_2}^{(-1,2)} = \frac{q_1 - q_2 }{384 \pi^4 R \mu_0 \pqty*{1- (R\mu_0)^2 }^3} \pqty*{H_1(R \mu_0) \left(q_1 - q_2 \right)^2 +q_1 q_2 H_2(R \mu_0, \alpha_{*}) },
\end{equation}
where
\begin{align}
  H_1 &
        \begin{multlined}[t][.88\textwidth]
          = - 9 (R\mu_0)^6-15 (R\mu_0)^4+45  (R\mu_0)^2+3 +4 \pi ^2 \left( (R\mu_0)^2-1\right)^2 \left(1-3  (R\mu_0)^2\right) \\+24 \pi  \left( (R\mu_0)^2-1\right) \sqrt{2  (R\mu_0)^2-3}  (R\mu_0)^2 \coth \left(\pi  \sqrt{2  (R\mu_0)^2-3}\right),
        \end{multlined} \\
  H_2 &\begin{multlined}[t][.7\textwidth]
    = 12 \left(1- (R\mu_0)^2\right) \bigg(4  (R\mu_0)^2(1 + (\pi -\alpha_*) \cot (\alpha_*))+(2 \pi -\alpha_*) \alpha_* \left(3  (R\mu_0)^4-4  (R\mu_0)^2+1\right) \\ -4 \pi   (R\mu_0)^2 \left(\csc (\alpha_*) \text{csch}\left(\pi  \sqrt{2  (R\mu_0)^2-3}\right) \sinh \left((\pi -\alpha_*) \sqrt{2  (R\mu_0)^2-3}\right) \right. \\ \left.+\sqrt{2  (R\mu_0)^2-3} \coth \left(\pi  \sqrt{2  (R\mu_0)^2-3}\right)\right)\bigg),
  \end{multlined}
\end{align}
with $\mu_0$ given by Eq.~\eqref{eq:mu0}.
Despite appearances, $\Gamma_{q_1 q_2}^{(-1,2)}$ is real and analytic for $R\mu_0 >  1$. By using Eq.~\eqref{eq:mu0} and performing expansions in the limits of small and large $\kappa$, one recovers Eq.~\eqref{eq:Gamma-12-expansion-small} and Eq.~\eqref{eq:Gamma-12-expansion-big} in the main text, respectively.

Finally, we present our result for the coefficient $ \beta_{q_1 q_2}$, which appears in the expansion of the cusp anomalous dimension about the straight-line limit $\alpha_* = \pi$ in Eq.~\eqref{eq:piexp}, expressed in terms of $\mu_0$:
\begin{align}
    \beta_{q_1q_2} & = Q \bigg[-\frac{G q_1 q_2}{12 \pi ^2}-\frac{G^2  (q_1-q_2) q_1 q_2 }{32 \pi ^4 R \mu_0  \left( (R\mu_0)  ^2-1\right)^2}\bigg(-3 (R\mu_0) ^4+\frac{16 (R\mu_0)  ^2}{3} \nonumber \\ & -\frac{4}{3} \pi  \left((R\mu_0) ^2-1\right) \sqrt{2  (R\mu_0)  ^2-3}  (R\mu_0) ^2 \text{csch}\left(\pi  \sqrt{2 (R\mu_0)  ^2-3}\right)-1\bigg)+ \order{G^3}\bigg] +\order{Q^0 G} \,.
\end{align}

\section{$\order{G}$ spectrum} \label{spettro}
Here we provide the eigenvalues of the inverse propagator for the fluctuations around the classical trajectory discussed in Sec.~\ref{symmental} to order $\order{G}$. We have
\begin{align}
      \lambda_{S,\pm}  ={}&-\omega ^2+J_\ell^2+\mu_0^2-m_d^2 \pm\sqrt{4 \mu_0^2 \omega ^2+\left(m_d^2-\mu_0^2\right)^2}\nonumber \\ &+G \left[ \frac{(q_1-q_2) \left(\sqrt{4 \mu_0^2 \omega ^2+\left(m_d^2-\mu_0^2\right)^2}\pm\left(m_d^2-5 \mu_0^2\right)\right)}{4 \pi ^2 R^3 \mu_0 \sqrt{4 \mu_0^2 \omega ^2+\left(m_d^2-\mu_0^2\right)^2}} \right. \nonumber \\ & \left. -\frac{\left(3 \sqrt{2} \pi  \left(m_d^2-\mu_0^2\right) \left(\sqrt{4 \mu_0^2 \omega ^2+\left(m_d^2-\mu_0^2\right)^2}\pm\left(\mu_0^2-m_d^2 \right)\right)\right)  I_r }{\sqrt{4 \mu_0^2 \omega ^2+\left(m_d^2-\mu_0^2\right)^2} \sqrt{\frac{q_1-q_2}{R \mu_0}}}\right. \nonumber \\ & \left. -\frac{\left(\mu_0 \left(-\mu_0^2+\sqrt{4 \mu_0^2 \omega ^2+\left(m_d^2-\mu_0^2\right)^2}+m_d^2+4 \omega ^2\right)\right) I_\mu}{\sqrt{4 \mu_0^2 \omega ^2+\left(m_d^2-\mu_0^2\right)^2}}  \right] \,, \\
         \lambda_{S,3}  ={}& -\omega ^2 +J_\ell^2 +\frac{q_1-q_2}{2 \pi ^2 R^3\mu_0}G \,, \\
         \lambda_{T}  ={}&-\omega ^2+J_{\ell,V}^2+(d-2)+\frac{q_1-q_2}{2 \pi ^2 R^3\mu_0}G \,, \\
         \lambda_{h}  ={}&-\omega ^2 +J_\ell^2 +\frac{q_1-q_2}{2 \pi ^2 R^3\mu_0}G  \,,\\
    \lambda_{\textrm{gh}} ={}&-\omega ^2 +J_\ell^2 +\frac{q_1-q_2}{2 \pi ^2 R^3\mu_0}G \,, \\
                \lambda_{II,1}  ={}&-\omega  (2 \mu_0+\omega )+J_\ell^2 +G \left[\sqrt{2} \pi  \left(\mu_0^2-m_d^2\right) \sqrt{\frac{R\mu_0}{q_1-q_2}} I_r-(\mu_0-2 \omega ) I_\mu\right] \,,\\
                  \lambda_{II,2}  ={}&\omega  (2 \mu_0+\omega )+J_\ell^2+G \left[\sqrt{2} \pi  \left(\mu_0^2-m_d^2\right) \sqrt{\frac{R \mu_0}{q_1-q_2}} I_r-(\mu_0+2 \omega ) I_\mu\right] \,,\\
 \end{align}
where 
\begin{align}
 I_r =  R \int Y_{\ell j m}^* Y_{\ell' j' m'}r_1  \dd \Omega_{d-1}    \,, \quad I_\mu = \int Y_{\ell j m}^*  Y_{\ell' j' m'}\mu_1 \dd \Omega_{d-1} \,,
\end{align}
and $J_{\ell,V}^2 =\frac{1}{R^2}(\ell (\ell+d-2)-1)$ denotes the eigenvalues of $\Delta_{S^{d-1}}$ acting on vectors which have multiplicity
\begin{equation}
  n_{\ell,V}= \frac{\ell (d+\ell-2) (d+2 \ell-2) \Gamma (d+\ell-3)}{\Gamma (d-2) \Gamma (\ell+2)} \,.
\end{equation}

The dispersion relations $\omega(\ell)$ are obtained by solving for zero eigenvalues in terms of $\omega$. Further setting $\ell=0$ yields the mass spectrum presented in the main text. Note that for $\ell=0$ one has $I_r=I_\mu = 0$ since the solutions satisfy $\int r_1 \dd  \Omega_{d-1} = \int \mu_1 \dd  \Omega_{d-1} = 0$.

\newpage
\printbibliography
\end{document}